\makeatletter\let\latex@xfloat\@xfloat\makeatother

\documentclass[numbers,sort&compress]{elsarticle}
\usepackage{etex}
\usepackage{lineno}
\usepackage[normalem]{ulem}
\usepackage{soul}
\usepackage[bookmarks,bookmarksopen,bookmarksdepth=2]{hyperref}
\hypersetup{colorlinks=true,citecolor=blue,linkcolor=blue}
\modulolinenumbers[5]

\usepackage{tikz}
\usetikzlibrary{positioning,decorations,arrows,patterns,shapes,backgrounds}
\usepackage{pgfplots}
\usetikzlibrary{shapes.misc}
\usetikzlibrary{patterns,decorations.pathreplacing}

\tikzset{cross/.style={cross out, draw=black, minimum size=2*(#1-\pgflinewidth), inner sep=0pt, outer sep=0pt},
cross/.default={1pt}}
\usepackage{gensymb}
\usepackage{indentfirst}
\usepackage{fancyhdr}
\usepackage{setspace}
\usepackage{dsfont}
\usepackage{amsfonts,amssymb,amsmath,mathtools,amsthm,ulem,cancel}
\usepackage{graphicx,color,geometry}
\usepackage{tikz}
\usepackage{pifont}
\usepackage{caption}
\usepackage{multirow}
\usepackage[export]{adjustbox}
\usepackage{subfig}
\usepackage{cleveref}
\usepackage{subfloat}

\graphicspath{{figures/}}

\definecolor{commenti}{rgb}{0.13,0.55,0.13}
\definecolor{stringhe}{rgb}{0.63,0.125,0.94}
\usepackage[font=footnotesize]{caption}

\usepackage{listings}
\usepackage{keyval}

\usepackage{tikz}
\usetikzlibrary{intersections}
\usetikzlibrary{calc,patterns,decorations.pathmorphing,decorations.markings}
\usepackage{pgfplots}
\usetikzlibrary{shapes.misc}
\usetikzlibrary{patterns,decorations.pathreplacing}

\tikzset{cross/.style={cross out, draw=black, minimum size=2*(#1-\pgflinewidth), inner sep=0pt, outer sep=0pt},
cross/.default={1pt}}

\newcommand{\numberset}{\mathds}
\newcommand{\R}{\numberset{R}}

\newcommand{\mb}{\mathbf}
\newcommand{\bs}{\boldsymbol}
\newcommand{\noi}{\noindent}

\definecolor{my-color}{RGB}{201,2,2}
\usepackage{soul}

\begin{document}

\begin{frontmatter}

\title{Numerical Simulations of Nearly Incompressible Viscoelastic Membranes}

\author[DMS]{Valeria Barra\corref{correspondingauthor}}
\cortext[correspondingauthor]{Corresponding author}
\ead{vb82@njit.edu}
\author[DMIE]{Shawn A. Chester}
\author[DMS]{Shahriar Afkhami}

\address[DMS]{Department of Mathematical Sciences,
New Jersey Institute of Technology,
Newark, NJ, 07102, USA}

\address[DMIE]{Department of Mechanical Engineering,
New Jersey Institute of Technology,
Newark, NJ, 07102, USA}

\begin{abstract}
This work presents a novel numerical investigation of the dynamics of free-boundary flows of viscoelastic liquid membranes. The governing equation describes the balance of linear momentum, in which the stresses include the viscoelastic response to deformations of Maxwell type. A penalty method is utilized to enforce near incompressibility of the viscoelastic media, in which the penalty constant is proportional to the viscosity of the fluid. A finite element method is used, in which the slender geometry representing the liquid membrane, is discretized by linear three-node triangular elements under plane stress conditions. Two applications of interest are considered for the numerical framework provided: shear flow, and extensional flow in drawing processes.
\end{abstract}
\begin{keyword}
Viscoelastic fluids; Membranes; Finite Elements
\end{keyword}

\end{frontmatter}

\section{Introduction}
Thin viscoelastic films can be found in a large variety of settings, from typical life situations to sophisticated manufacturing processes. In our everyday life, we may encounter sheets or thin layers of liquids that show a viscoelastic behavior, such as custard, shampoo, shaving cream, wax, glue, and paint; or similarly, soft solids with the same characteristics, such as gels. For biomedical engineering applications, thin viscoelastic sheets can represent biopolymers \cite{LevineMacKintosh}, or biological tissues constituting blood cells \cite{EvansHochmuth,LubardaMarzani}. In some manufacturing processes, thin layers of elastic or viscoelastic materials, for instance, in the form of liquid crystal polymers, are largely employed \cite{Callister}. Hence, the prediction of the behavior of viscoelastic sheets through mathematical and numerical modeling becomes a cost-effective manufacturing practice, as well as an important tool to better understand some physical effects, that are difficult or expensive to reproduce experimentally.
The mathematical and numerical framework developed in this work aims at providing insight to the understanding of the dynamics and physical behavior of thin layers of viscoelastic media, modeled as membranes.

Thin curved bodies are commonly modeled as shells or membranes \cite{GZ,TaylorEtAl2005}. The slender geometry of thin films or sheets of various materials can be described through an idealized mid-surface, that sits at half thickness between the top and bottom surfaces of the sheet. For the general theory of shells, the mid-surface has a non zero curvature, and any application of loading or external forces causes both bending and stretching \cite{Ribe2,Howell}. A particular case of this general theory is the membrane theory of shells, that concerns the study of the in-plane stretching deformations, dominant with respect to transversal deflections, and in which bending stiffness is neglected. In this work, we utilize the membrane theory of shells, in which the in-plane stresses are included to model the viscoelastic response to deformations. The majority of the studies in the literature of membrane theory of shells, focuses on the statics of load-carrying elastic shells that hold an equilibrium state (see, for instance \cite{Love,GZ,LandauLifshitz,VentzelKrauthammer,BonetEtAl}). However, in this work we are interested in the transient analysis of the dynamics, described by the conservation of momentum equation, as outlined by Taylor et al.~in \cite{TaylorEtAl2005}, for the case of nonlinearly elastic membranes. Our goal is to expand the analysis conducted by Taylor et al.~to include Newtonian and non-{N}ewtonian membranes. For the non-{N}ewtonian membranes we characterize the stresses by the Maxwell model \cite{Maxwell}. We use this infinitesimal strain model within the general framework developed by Taylor et al.~in \cite{TaylorEtAl2005} for finite strain theory, with the aim of expanding our analysis in future works, by including nonlinearities and corotational effects.

Viscoelastic materials exhibit features that are typical of both fluids (viscosity) and solids (elasticity). This hybrid nature allows it to characterize a broad variety of materials, with limiting cases that fall under a liquid state, or a solid state, and intermediate regimes that constitute soft materials, such as gels \cite{Chester}. The evolution of their complex internal microstructures can affect their dynamics and the overall macroscopic rheology \cite{YueFengLiuShen}. The majority of the previous studies on viscoelastic membranes focus on the rheological responses of the material to deformations (see, e.g.,~\cite{EvansHochmuth,ChuaOyen,HarlandEtAl}), but only a few works investigate the dynamics of such membranes; see, for instance, \cite{LevineMacKintosh}, in which the dynamics of the viscoelastic membrane is coupled to the hydrodynamics of the surrounding viscous phase. Among the numerous studies on the rheology of viscoelastic membranes, Lubarda and Marzani \cite{LubardaMarzani} use the Kelvin-Voigt constitutive model, that is more suitable to describe viscoelastic solids \cite{MainardiSpada}; while Crawford and Earnshaw \cite{CrawfordEarnshaw} use the Maxwell model, more suitable for the description of viscoelastic liquids \cite{Bird}, to identify the relaxation time of bilayer lipid membranes. Moreover, some studies propose numerical solutions of the dynamics of thin layers of viscoelastic fluids within the lubrication theory to simulate the interfacial flow of thin viscoelastic films of Jeffreys type \cite{Jeffreys}, deposited on substrates, in wetting or dewetting processes \cite{VilminRaphael,TomarEtAl,BarraEtAl}. However, to the best of our knowledge, a numerical investigation solving for the equation of motion describing the hydrodynamics of the free-boundary flow of thin viscoelastic membranes of Maxwell type is not available in the literature. The aim of this work is therefore to provide a general numerical framework for the simulations of thin viscoelastic membranes, and to analyze the role of viscoelasticity on their dynamics arising in different settings or engineering processes, such as shearing flows \cite{Bird} or stretching in redraw processes \cite{oKielyBrewardGriffiths2015,Taroni2013}.

The Maxwell model belongs to a class of linear differential models for non-{N}ewtonian fluids, that describes mechanical properties such as ``fading memory'' and stress relaxation \cite{Bird}. These features become remarkable, especially when compared to constitutive models that describe a linear relationship between the stress and the strain (for linear elastic solids) or strain rate (for Newtonian fluids). The Maxwell constitutive model, in the same fashion as Hooke's law, was proposed empirically \cite{Maxwell}. Although it has been applied to and proven to be useful for the analysis of a broad range of materials, this model is limited to cases in which the deformation gradients are infinitesimally small \cite{Bird}. To overcome this limitation, variations of Maxwell model have been proposed, such as the Oldroyd-B model \cite{Bird} in which convective derivatives are introduced to describe the nonlinearities in the stress tensor. Despite the limitations of a linear viscoelastic model, such as the Maxwell model, we believe that a comprehensive analysis as well as a detailed numerical framework for the dynamics of thin viscoelastic membranes, can serve as a benchmark for future analyses that include nonlinear features, such as the convective/corotational variations of the stress.

The governing equation describes the conservation of linear momentum. To the typical steady formulation in which the balance of forces is considered, we retain the inertial term so we may consider transient analyses \cite{ZienkiewiczTaylorZhu2013,Palaniappan}. The incompressibility condition that typically serves as a constraint on the vector velocity field in the equations describing the fluid dynamics \cite{Batchelor} is replaced in this work via the use of a \textit{penalty method} \cite{Hughes,VanDerZanden}. This method, first introduced by Courant \cite{Courant} for solutions of problems of equilibrium and vibrations, obtained by the calculus of variations, has been subsequently used to approximate solutions of the Navier-Stokes equations (see, for instance, \cite{Shen} and the references therein). In the context of solutions of fluid flows, it relaxes the incompressibility condition allowing for a small perturbation of the rate of volume change, which approximates the near incompressibility of the fluid. We propose a formulation of the penalty function as a direct proportionality on the rate of change of the volumetric strain, in which the constant of proportionality depends on the viscosity of the fluid.

In this numerical investigation, we use the finite element method for the spatial discretization of the slender geometry describing the membranes, and implicit schemes to discretize the time variations in the governing and constitutive equations. Finite element analyses of linearly elastic shells or membranes constitute a computational advantage relative to volumetric analyses and are vast in the Continuum Mechanics literature (see, e.g.~\cite{ZienkiewiczTaylorFEM2000,TaylorEtAl2005,ZienkiewiczTaylorZhu2013,HansboEtAl2015}), but, to the best of our knowledge, none of the existing analyses included viscoelastic stresses of Maxwell type that can be adapted to Fluid Mechanics problems. We approximate the membrane with a mesh, constituted of linear $3$-node triangles embedded in a three-dimensional global coordinate system (i.e.~elements with nine degrees of freedom with respect to the global coordinates), and obtain the stress state on the surface of the membrane in terms of the nodal displacements. The spatial discretization formulation adopted closely follows the one by Taylor et al.~\cite{TaylorEtAl2005}, however, the novel aspects are the inclusion of viscoelasticity in the constitutive model, the corresponding derivation of the material Jacobian (stiffness) tensor, and the numerical investigation of the dynamics of viscoelastic liquid membranes in free-boundary, shear and extensional flows.

The remainder of this work is organized as follows: In \S~\ref{Sec2}, we introduce the mathematical formulation and finite element analysis of the governing equation (whose detailed theoretical derivation is given in the Appendix); In \S~\ref{Sec3}, we introduce the material models considered in this analysis both in continuous and discrete form; In \S~\ref{Sec4}, we discuss our numerical results; In \S~\ref{Sec5}, we draw our conclusions.

\section[Mathematical Formulation]{Mathematical Formulation}\label{Sec2}
We consider a nearly incompressible viscoelastic liquid membrane with constant density $\rho$, surrounded by a passive gas with constant pressure. The equation describing the balance of linear momentum is
\begin{align}\label{Momentum}
\textrm{div}( \boldsymbol{\sigma}) + \mathbf{F_b} &= \rho \mathbf{\ddot{u}},  \qquad \textrm{ in } \Omega\, ,
\end{align}

\noi where $\textbf{u} = (u_1(x_1,x_2,x_3,t),u_2(x_1,x_2,x_3,t),u_3(x_1,x_2,x_3,t))$ represents the vector displacement field in a global coordinate system, $\mathbf{\ddot{u}} = d^2 \mathbf{u} / d t^2$ in a Lagrangian formulation, $\mathbf{F_b}$ is the vector of the body force (such as gravity), $\textrm{div}( \bs{\sigma}) = \nabla \cdot \bs{\sigma}$, with $\bs{\sigma}$ the symmetric stress tensor, and $\Omega$ is the two-dimensional surface embedded in $\R^3$. In what follows, we outline the weak and discrete versions of equation (\ref{Momentum}), leaving the detailed derivation for the interested reader in the Appendix.

\begin{figure}[t]
\centering
\resizebox{.45\textwidth}{!}{
\begin{tikzpicture}
\draw[->] (0,0)--(1,0) node[anchor=west] {$x_2$};
\draw[->] (0,0)--(0,1) node[anchor=east] {$x_3$};
\draw[->] (0,0)--(-.75,-.75) node[anchor=east] {$x_1$};
\draw (.9,.9) -- (2.25,1.25)node[anchor=north] {2};
\draw[->] (2.25,1.25) -- (2.9,1.411)node[anchor=west] {$y_1$};
\draw[->] (.9,.9) -- (.66,2.247)node[anchor=east] {$y_2$};
\draw[->,>=stealth] (.9,.9) -- (1.2,1.2);
\draw(1.1,1.15)node[anchor= south west] {$\mathbf{n}$};
\draw (2.25,1.25) -- (1.35,2.25)node[anchor=south] {3};
\draw (1.35,2.25) -- (.9,.9)node[anchor=north] {1};
\end{tikzpicture}%
}
\caption{The surface coordinate system on a triangular element in the deformed configuration.}\label{fig:SurfCoordSys}
\end{figure}
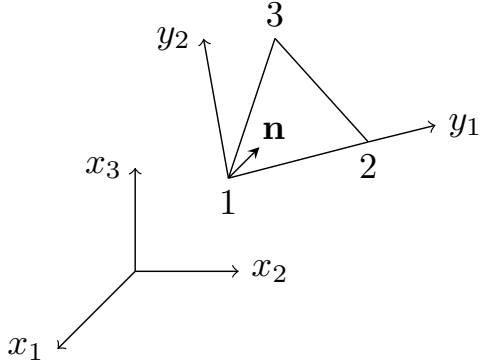

We discretize the domain $\Omega$ with finite elements, in which each element represents a triangular membrane under plane stress conditions, uniquely described by its three vertices (nodes) in $\R^3$ (see figure \ref{fig:SurfCoordSys}). By considering a global Cartesian coordinate system, we denote by upper case $\mathbf{X}$ the reference (undeformed state) configuration coordinates, and by lower case $\mathbf{x}$ the current (deformed state) ones. We denote the nodal values of the reference coordinates, current coordinates and displacement vector, respectively, by the use of superscripts, i.e~$\widetilde{\mathbf{X}}^\alpha$, $\widetilde{\mathbf{x}}^\alpha$, and $\widetilde{\mathbf{u}}^\alpha = \widetilde{\mathbf{x}}^\alpha - \widetilde{\mathbf{X}}^\alpha$, with $\alpha = 1,2,3$ for each node. By using the virtual displacement field, $\delta \mb{u}$, we apply the virtual work formulation \cite{TaylorEtAl2005,ZienkiewiczTaylorZhu2013}, and obtain the weak form of (\ref{Momentum}) as
\begin{align}\label{Eq:25}
\delta \Pi = \int_{\Omega^{(e)}}\delta \mb{u}^T  \rho \mb{\ddot{{u}}} \,d\, V -\int_{\Omega^{(e)}}  \delta \bs{\epsilon} ^T \bs{\sigma} \,d\, V -\int_{\Omega^{(e)}}\delta \mb{u}^T \mb{F_b}  \,d\, V \,= \mb{0} \, ,
\end{align}

\noi where $[\cdot]^T$ represents the matrix transpose operator, $\bs{\epsilon}$ the symmetric strain tensor, and $\Omega^{(e)}$ the domain of the element $e$. For the case of membranes of constant thickness $h$, we express an infinitesimal volume element as $dV = h \,dA$. Following the displacement-based finite element formulation provided in \cite{TaylorEtAl2005} for the spatial derivatives (outlined in the Appendix), we can write the spatially discrete version of the volume contribution terms (i.e.~without the traction term) of equation (\ref{Eq:25}), for each element, in vector form, as
\begin{align}\label{Eq:41}
\mb{M}^{(e)}
\left[
\begin{array}{c}
\mb{\ddot{\tilde{u}}}^1\\
\mb{\ddot{\tilde{u}}}^2\\
\mb{\ddot{\tilde{u}}}^3
\end{array}
\right]
- h A^{(e)} \mb{B}^{(e)T}
\left[
\begin{array}{c}
\sigma_{11}\\
\sigma_{22}\\
\sigma_{12}
\end{array}
\right]
- \left[
\begin{array}{c}
\mb{\widetilde{F}_b}^1\\
\mb{\widetilde{F}_b}^2\\
\mb{\widetilde{F}_b}^3
\end{array}
\right] = \mb{0}\, ,
\end{align}

\noi where we have used Voigt notation \cite{ZienkiewiczTaylorZhu2013} for the symmetric stress tensor in vector form for two-dimensional problems, defined by
\begin{align*}
\bs{\sigma} = \left[
\begin{array}{c}
\sigma_{11}\\
\sigma_{22}\\
\sigma_{12}
\end{array}\right] \, ,
\end{align*}

\noi and where $A^{(e)}$ represents the area of each triangular element in the reference configuration; the vector $\mb{\widetilde{F}_b} = (\mb{\widetilde{F}_b}^1,\mb{\widetilde{F}_b}^2,\mb{\widetilde{F}_b}^3)$ represents the nodal body force; $\mb{M}^{(e)}$ is the element mass matrix, and $\mathbf{B}^{(e)T} \bs{\sigma}$ represents the divergence of the stress tensor on each element. In each triangle, we consider that both the strain and the stress tensors are constant. The interested reader can find the details of the derivation of each term of equation (\ref{Eq:41}) in the Appendix. Our goal is to solve equation (\ref{Eq:41}) for the nodal displacement field. We note that the nodal displacement vectors, $\mb{\tilde{u}}^\alpha$ ($\alpha = 1,2,3$), as well as the nodal force vectors, $\mb{\widetilde{F}_b}^{\alpha}$ ($\alpha = 1,2,3$), represent three-dimensional vectors for each node, in the global coordinates. Hence, in components, we will solve for nine scalar equations, even though the strain and the stress tensors only account for the in-plane displacements.

\section[Constitutive Models]{Constitutive Models}\label{Sec3}
To describe the material response to deformations, we need to express a constitutive law that relates the stress tensor and the strain and/or strain rate tensors. We consider a small deformation strain, within the general framework presented by Taylor et al.~\cite{TaylorEtAl2005} that allows nonlinearities due to large deformations (derived in the Appendix). For membrane problems, the in-plane magnitudes of the stress are dominant relative to the out-of-plane ones, leading to the conditions (referred to as plane stress conditions) on the stress tensor components, $\sigma_{13}=\sigma_{23}=\sigma_{33}=0$. In two spatial dimensions, the \emph{deviatoric} stress is defined, in tensor form, as
\begin{align}\label{DeviatoricStress}
{\sigma'_{ij}} = \sigma_{ij} - \frac{1}{2}  \sigma_{kk}\delta_{ij} \, ,
\end{align}

\noi where $\delta_{ij}$ is the Kronecker delta ($i,j=1,2$), and $\sigma_{kk}$ is the trace of the stress tensor in indicial notation, i.e.~$\sigma_{kk}=\sigma_{11} +\sigma_{22}$. In infinitesimal strain theory \cite{ZienkiewiczTaylorZhu2013}, the linear (small deformation) strain is given, in tensor form, by $ \epsilon_{ij} = \left( {\partial u_i}/{\partial x_j}+ {\partial u_j}/{\partial x_i} \right) / 2$. In two dimensions, the trace of the strain tensor, also called the volumetric strain, is denoted by $\epsilon_{vol} = \epsilon_{kk} = \epsilon_{11} +\epsilon_{22}$. We call hydrostatic strain the mean of the normal strains, that is, $\epsilon_{hyd} = \epsilon_{kk}  / 2$. With this definition, we can also define the deviatoric strain, $\epsilon_{ij}'$, satisfying
\begin{align}\label{Eq:Strain}
\epsilon_{ij}' = \epsilon_{ij} - \frac{1}{2}  \epsilon_{kk} \delta_{ij} \, .
\end{align}

\noi An important material parameter related to the response to (uniform) pressure in linear elasticity of isotropic media is the bulk modulus, $K$, and it is related to other material parameters such as $\nu$, the Poisson's ratio, and $Y$, the Young's modulus, via the relationship $\nu ={1}/{2} - {Y}/{6K}$ \cite{ZienkiewiczTaylorZhu2013}. We notice that for $K \gg Y$, meaning in the limiting case in which $K \rightarrow \infty$ (i.e.~for $\nu \rightarrow 1/2$), we approach the incompressible limit. However, for nearly incompressible materials, a \textit{penalty function} \cite{Hughes} that allows for small perturbations to the trace of the strain, representing the volumetric change, is given by
\begin{align}\label{VolStrain}
\epsilon_{kk} + p/K = 0 \, .
\end{align}

\noi Hence, we find an expression for the pressure, $p$, in terms of the volume variation, given by
\begin{align}\label{HydrostaticPressure}
p = -{K} {\epsilon}_{kk} \, .
\end{align}


\noi For various formulations, including the penalty method, the reader is referred to \cite{ZienkiewiczTaylorZhu2013,Hughes}.

In this work, we expand the condition (\ref{VolStrain}) to account for the hydrodynamic pressure, $p$, in liquids. In constitutive models describing liquids, the stress response is directly proportional not to the strain, but to the rate of change of the strain, namely $ \dot{\bs{\epsilon}}$. Accordingly, the consideration of a penalty method for liquids needs to take into account the strain rate \cite{VanDerZanden}. We introduce a penalty formulation for the variation of the volume of nearly incompressible liquids
\begin{align}\label{VolStrainRate}
\dot{\epsilon}_{kk} + p/ \widehat{K} = 0  \, ,
\end{align}

\noi for which the pressure in the liquid is then given in terms of the trace of the strain rate by
\begin{align}\label{Pressure}
p= -\widehat{K} \dot{\epsilon}_{kk} \, ,
\end{align}

\noi with the penalty constant, $\widehat{K}$, such that
\begin{align}\label{KHat}
\widehat{K} \gg \eta \, ,
\end{align}

\noi where $\eta$ represents the shear viscosity coefficient.

We start our constitutive analysis by introducing the Newtonian constitutive model for viscous liquids, given by
\begin{align}\label{Newtonian}
{\sigma}_{ij}= 2\eta \dot{\epsilon}_{ij}' + \widehat{K}\dot{\epsilon}_{kk}\delta_{ij}  \, .
\end{align}

\noi Next, we include in our analysis viscoelastic fluids. Different linear viscoelastic constitutive models of interest can be expressed in linear differential form \cite{Bird,MainardiSpada,Siginer}. The Maxwell constitutive model for viscoelastic liquids is given by
\begin{align}\label{Maxwell}
{\sigma}_{ij} + \tau \partial_t {\sigma}'_{ij}= 2\eta \dot{ \epsilon}_{ij}' + \widehat{K}\dot{\epsilon}_{kk}\delta_{ij}  \, ,
\end{align}

\noi where $\tau$ is the relaxation time constant, such that $\tau =  \eta / G$, with $G$ the shear modulus \cite{Bird}. We notice that when $\tau=0$ we recover the Newtonian fluid constitutive law in equation (\ref{Newtonian}). When $\tau > 0$, it determines the rate at which the stress relaxes (i.e~decays) for constant strain. Maxwell model can interpolate between a linearly viscous and elastic behavior. In fact, when the stress applied has a fast time variation, the left hand side of equation (\ref{Maxwell}) is dominated by the time derivative, and, upon time integration, the constitutive law for linearly elastic solids is recovered \cite{Bird}.

\subsection[Time Discretization]{Time Discretization}\label{Sec3-1}
The time interval $t \in [0,T]$ is discretized by $n$ equal steps, with $n=0,1, \ldots$, and $\Delta t$ is the temporal step size considered. At each spatial material point, we denote the stress at the previously converged time step by $\bs{\sigma}^{n}$, and at the current time step by $\bs{\sigma}^{n+1}$. We define the rate of change of the strain tensor with a finite difference $\bs{\dot{\epsilon}}^{n+1} = (\bs{\epsilon}^{n+1} - \bs{\epsilon}^n)/\Delta t$, and similarly for the stress tensor. We consider initial conditions on both the strain and the stress to be $\bs{\epsilon}^0 = \bs{\sigma}^0 = \mathbf{0}$. Hence, we can write the discrete form of equation (\ref{Newtonian}), in indicial form, as
\begin{align}\label{DiscreteNewtonianTensor}
{\sigma}_{ij}^{n+1}= \frac{2\eta}{\Delta t} \left\{ \left( {\epsilon}_{ij} - \frac{1}{2}{\epsilon}_{kk}\delta_{ij}\right)^{n+1} - \left( {\epsilon}_{ij} - \frac{1}{2}{\epsilon}_{kk}\delta_{ij}\right)^{n} \right\} + \frac{\widehat{K}}{\Delta t}\left\{ {\epsilon}_{kk}^{n+1}\delta_{ij} - {\epsilon}_{kk}^{n}\delta_{ij}\right\}  \, ,
\end{align}

\noi and in vector form as
\begin{align}\label{DiscreteNewtonianVector}
\left[
\begin{array}{c}
\sigma_{11}\\
\sigma_{22}\\
\sigma_{12}
\end{array}
\right]^{n+1}=&
\frac{2\eta}{\Delta t}\left\{
\left[
\begin{array}{c}
\frac{1}{2}\epsilon_{11} - \frac{1}{2} \epsilon_{22}\\
- \frac{1}{2} \epsilon_{11} + \frac{1}{2}\epsilon_{22}\\
\gamma_{12}
\end{array}
\right]^{n+1} -
\left[
\begin{array}{c}
\frac{1}{2}\epsilon_{11} - \frac{1}{2} \epsilon_{22}\\
- \frac{1}{2} \epsilon_{11} + \frac{1}{2}\epsilon_{22}\\
\gamma_{12}
\end{array}
\right]^{n} \right\}
+ \nonumber \\
& \frac{\widehat{K}}{\Delta t}\left\{
\left[
\begin{array}{c}
\epsilon_{11} + \epsilon_{22} \\
\epsilon_{11} + \epsilon_{22} \\
0
\end{array}
\right]^{n+1} -
\left[
\begin{array}{c}
\epsilon_{11} + \epsilon_{22} \\
\epsilon_{11} + \epsilon_{22} \\
0
\end{array}
\right]^{n}\right\}\, ,
\end{align}

\noi where we have used the notation $\gamma_{12} = 2 \epsilon_{12}$, for which in vector form shear strain components are twice that given in tensor form \cite{ZienkiewiczTaylorZhu2013}. We consider the algorithmic consistent Jacobian (or stiffness) fourth order tensor to be defined for the case of Newtonian fluids as
\begin{align}\label{CTensorNewtonian}
\mathbb{C}_{ijlm} = \frac{\partial \sigma_{ij}^{n+1}}{\partial \dot{\epsilon}^{n+1}_{lm}} = \eta (\delta_{il}\delta_{jm} + \delta_{im} \delta_{jl}) -  \eta\delta_{ij}\delta_{lm} + \widehat{K}\delta_{ij}\delta_{lm} \,.
\end{align}

\noi This way, we can write equation (\ref{Newtonian}) in matrix form as a linear relation between the stress tensor and the strain rate tensor with a constant coefficient matrix, $\mathbf{D}_v$. We shall refer to $\mathbf{D}_v$ as the \textit{viscosity matrix of moduli}, analogously to the elasticity matrix of moduli \cite{ZienkiewiczTaylorZhu2013}, and write
\begin{align}\label{NewtonianElasticModuli}
\bs{\sigma}^{n+1} = \mathbf{D}_v \dot{ \bs{\epsilon}}'^{n+1} \, ,
\end{align}

\noi in components,
\begin{align}\label{NewtonianCTensorCoponents}
\mathbf{D}_v=
\left[
\begin{array}{ccc}
\frac{c_v}{2} + \widehat{K}  &
-\frac{c_v}{2} + \widehat{K}&
0 \\
\\
-\frac{c_v}{2} + \widehat{K}&
\frac{c_v}{2} + \widehat{K} &
0 \\
\\
0 &
0 &
\frac{c_v}{2}
\end{array}
\right]\, ,
\end{align}

\noi where we have used the constant $c_v=2 \eta $.

Similarly, we discretize the Maxwell model, in equation (\ref{Maxwell}), by considering the stress implicitly. In indicial form, it becomes
\begin{align}\label{DiscreteMaxwellTensor}
{\sigma_{ij}'}^{n+1} =&\left( 1 + \frac{\Delta t}{\tau}  \right)^{-1}  \left\{ {\sigma_{ij}'}^n  + \frac{2 \eta}{\tau}\left[\left({\epsilon}_{ij} - \frac{1}{2}{\epsilon}_{kk}\delta_{ij}\right)^{n+1} - \left({\epsilon}_{ij} - \frac{1}{2}{\epsilon}_{kk}\delta_{ij}\right)^{n}\right] +\right. \nonumber \\
& \left. \frac{\widehat{K}}{\tau}\left[{\epsilon}_{kk}^{n+1}\delta_{ij} - {\epsilon}_{kk}^{n}\delta_{ij}\right]\right\}\, ,
\end{align}

\noi which may be written in vector form, as
\begin{align}\label{DiscreteMaxwellVector}
\left[
\begin{array}{c}
\sigma_{11}'\\
\sigma_{22}'\\
\sigma_{12}'
\end{array}
\right]^{n+1}=&
\left( 1 + \frac{\Delta t}{\tau}  \right)^{-1}
\left\{
\left[\begin{array}{c}
\sigma_{11}'\\
\sigma_{22}'\\
\sigma_{12}'
\end{array}\right]'^n  +
\frac{2 \eta}{\tau} \left[
\left[\begin{array}{c}
\frac{1}{2}\epsilon_{11} - \frac{1}{2} \epsilon_{22}\\
- \frac{1}{2} \epsilon_{11} + \frac{1}{2}\epsilon_{22}\\
\gamma_{12}\end{array}\right]^{n+1}
-
\left[\begin{array}{c}
\frac{1}{2}\epsilon_{11} - \frac{1}{2} \epsilon_{22}\\
- \frac{1}{2} \epsilon_{11} + \frac{1}{2}\epsilon_{22}\\
\gamma_{12}\end{array}\right]^{n}
\right]\right.+\nonumber \\
& \left.  \frac{\widehat{K}}{\tau}
\left[
\left[
\begin{array}{c}
\epsilon_{11} + \epsilon_{22} \\
\epsilon_{11} + \epsilon_{22} \\
0
\end{array}
\right]^{n+1} -
\left[
\begin{array}{c}
\epsilon_{11} + \epsilon_{22} \\
\epsilon_{11} + \epsilon_{22} \\
0
\end{array}
\right]^{n}\right]
\right\}\, .
\end{align}

\noi Now we can rewrite the relation in equation (\ref{Maxwell}) in matrix form, with a constant coefficient matrix, $\mathbf{D}_{ve}$, that we shall call the \textit{viscoelasticity matrix of moduli}. This matrix does not express a direct proportionality between the variation of the stress and the one of the strain, as in the viscous case. But it expresses the variation of the total change of the algorithmic stress (including its history) with respect to the rate of change of the strain, that is
\begin{align}\label{ViscoeaslticElasticModuli}
\bs{\sigma}^{n+1} + \tau \partial_t{\bs{\sigma}}'^{n+1} = \mathbf{D}_{ve} \dot{\bs{\epsilon}}'^{n+1} \, ,
\end{align}

\noi where in components $\mathbf{D}_{ve}$ has the same form as the Newtonian one, in equation (\ref{NewtonianCTensorCoponents}), except for the constant that now is defined as $c_{ve}= c_{v}/\tau$, and $\widehat{K}/\tau$ appears in place of $\widehat{K}$.

The discrete material models presented are implemented as a user defined subroutine (\emph{UMAT}) for the software Abaqus/Standard $6.13$, and the time derivatives of equation (\ref{Eq:41}) are discretized implicitly with a generalized Newmark scheme \cite{HilberHughesTaylor}.

\begin{figure}[t]
\centering
\centering
\resizebox{0.35\textwidth}{!}{
\begin{tikzpicture}
\draw (0,0) -- (2.5,0);
\draw (0,-0.2) -- (0,-0.1)node[anchor=north] {};
\draw (0,-0.15) -- (2.5,-0.15)node[anchor=north] {};
\draw (2.5,-0.2) -- (2.5,-0.1)node[anchor=north] {};
\draw (1.25,-0.25) node[anchor=north]{$44\, $mm};
\draw (0,0) -- (2.5,3); 
\draw (2.5,3) -- (3.5,3); 
\draw (2.5,-0.15) -- (3.5,-0.15)node[anchor=south] {};
\draw (2.5,-0.2) -- (2.5,-0.1)node[anchor=south] {};
\draw (3.5,-0.2) -- (3.5,-0.1)node[anchor=south] {};
\draw (3,-0.25) node[anchor=north]{$16\, $mm};
\draw[dashed] (2.5,0) -- (2.5,3) ; 
\draw (3.35,3.375) node[anchor=east]{$\mb{P}$};
\draw[->] (2.5,3.075)--(4,3.075) node[anchor=east] {};
\draw (3.65,0) -- (3.65,3)node[anchor=west] {};
\draw (3.6,0) -- (3.7,0)node[anchor=west] {};
\draw (3.6,3) -- (3.7,3)node[anchor=west] {};
\draw (3.65,1.5) node[anchor=west]{$48\, $mm};
\draw (3.5,3) -- (2.5,0);
\end{tikzpicture}%
}

\caption{Cook's membrane schematic for the numerical experiment.}\label{fig:Cook}%
\end{figure}
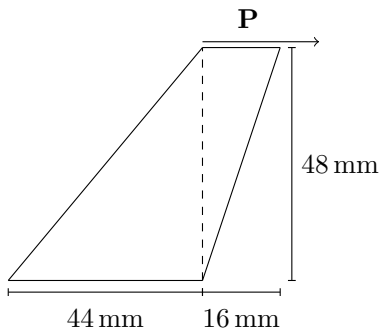

\begin{figure}[t]
\subfloat[]{\includegraphics[width=8cm,valign=t,trim=0.15in 0.15in 0.2in 0.15in,clip=true]{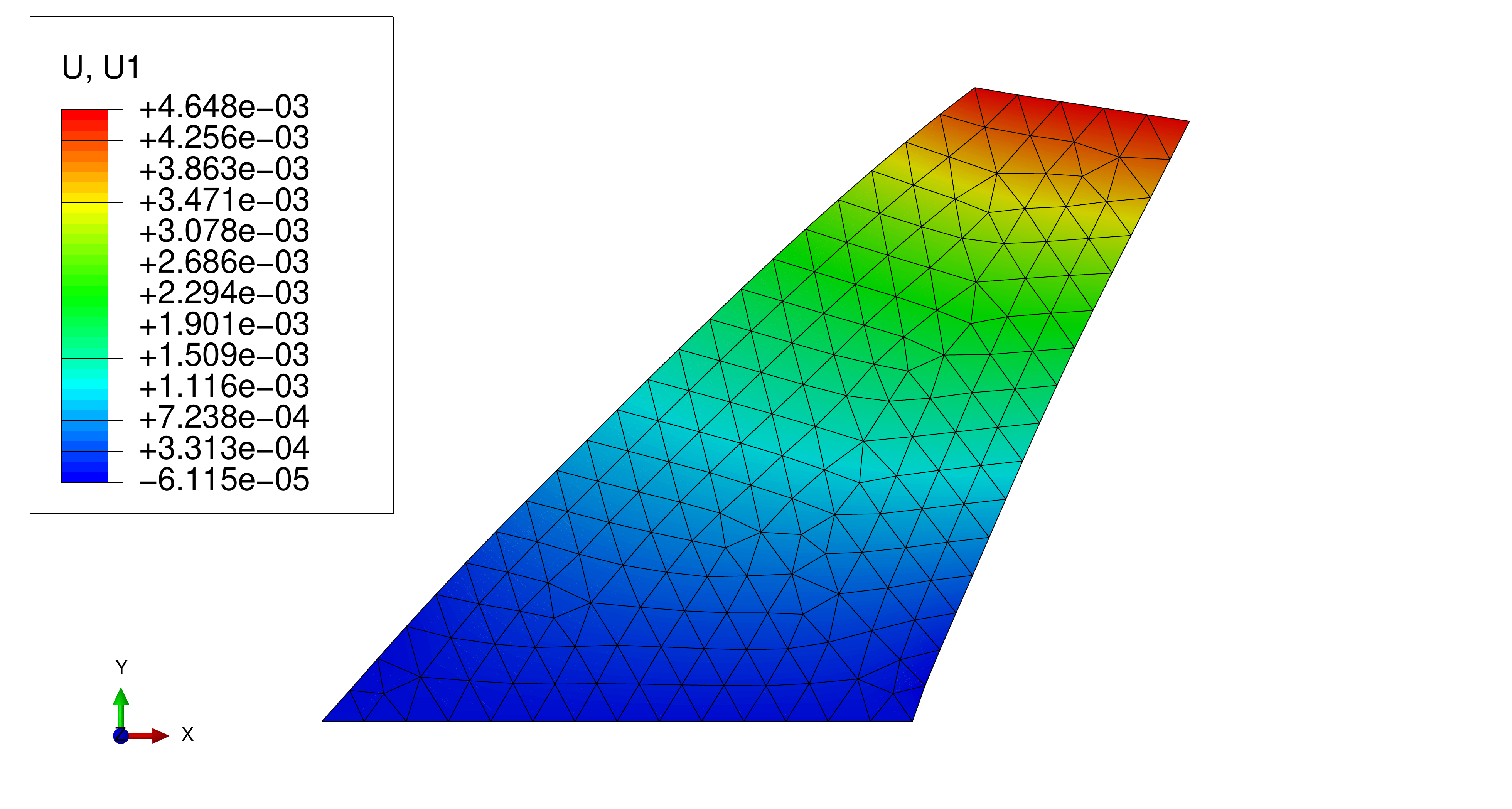}\label{fig:CookLastStepU1}}
\subfloat[]{\includegraphics[width=8cm,valign=t,trim=0.15in 0.15in 0.2in 0.15in,clip=true]{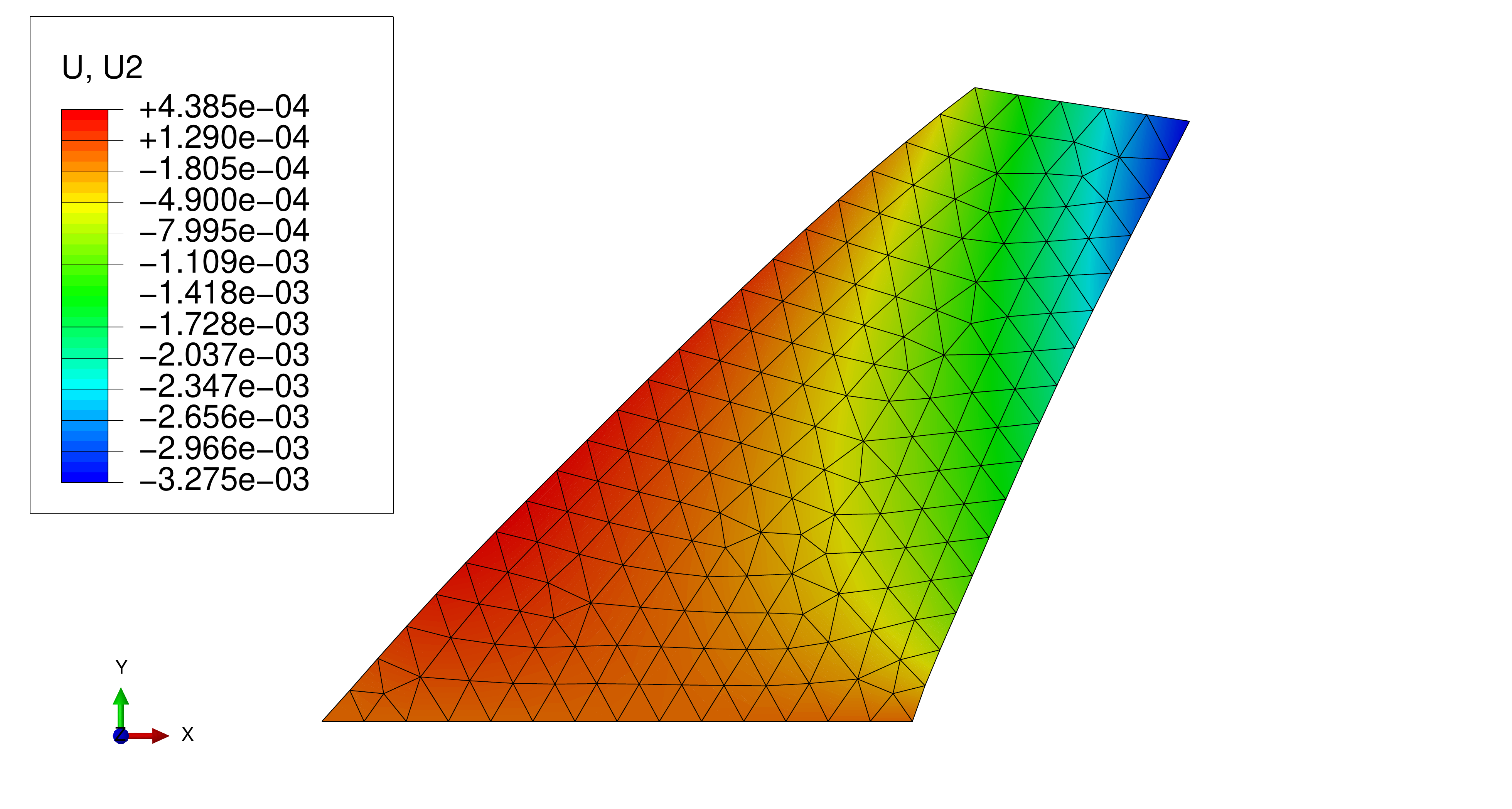}\label{fig:CookLastStepU2}}
\caption{The final configuration, at $t^{\star}=1$, of a Cook's membrane of viscoelastic material of Maxwell type, with viscosity coefficient $\eta = 10 \, Pa \, s$, and relaxation time $\tau = 1 \, s$. The color gradient represents contour plots of the displacement field, in which warmer shades mean higher values. In \protect\subref{fig:CookLastStepU1}, we show the first component of the displacement field, $u_1$, that ranges between its minimum value, ${u_1}_{min} \sim0$~m (blue), and its maximum value, ${u_1}_{max}=4.648 \times 10^{-3}$~m (red). In \protect\subref{fig:CookLastStepU2}, we display the second component, $u_2$, that ranges between its minimum value, ${u_2}_{min}=-3.275 \times 10^{-3}$~m (blue), and its maximum value, ${u_2}_{max}=4.385\times10^{-4}$~m (red).}\label{fig:CookLastStep}
\end{figure}

\begin{figure}[t]
\centering
{\includegraphics[height=5cm,valign=t,clip=true]{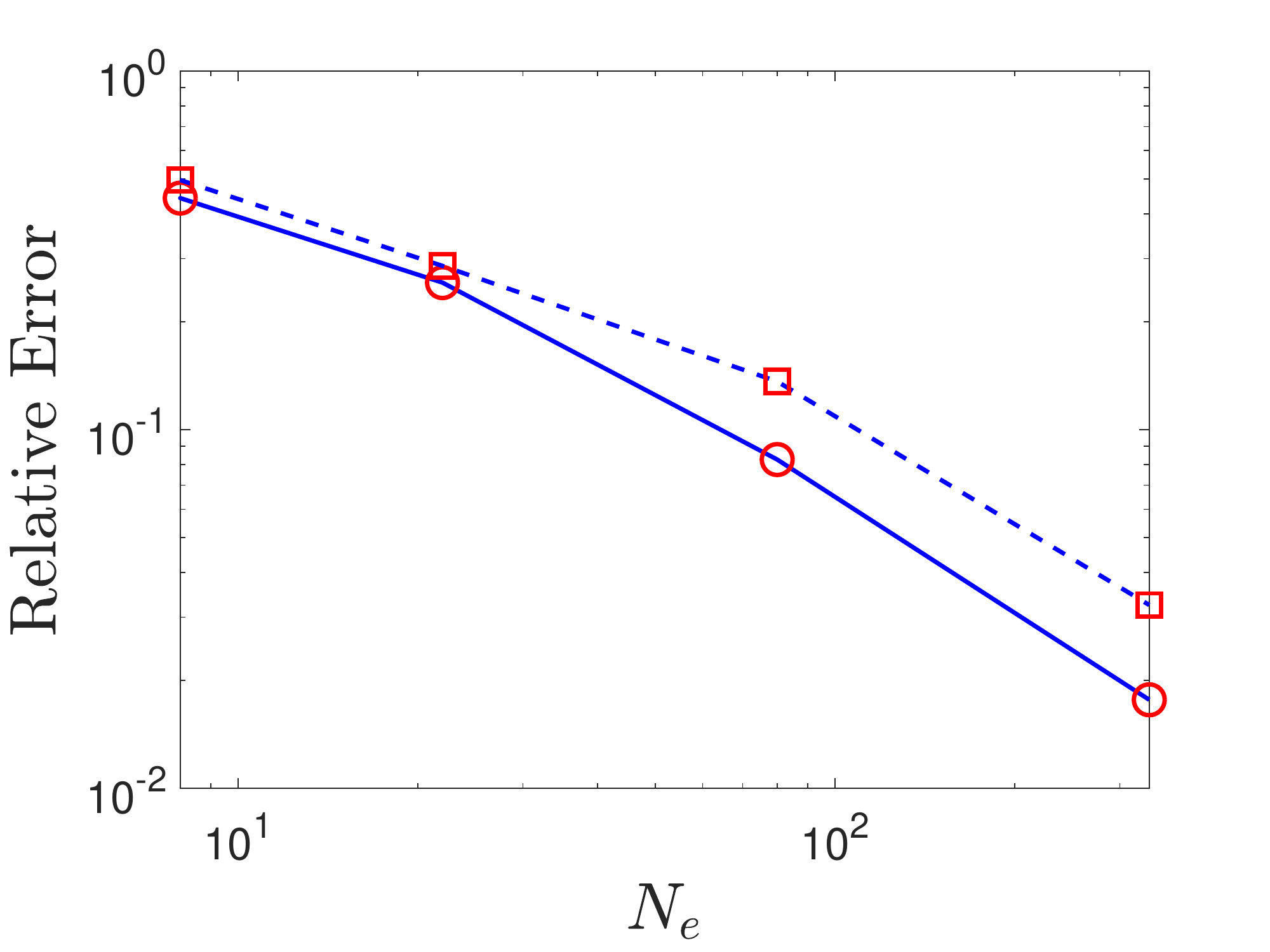}}%
\caption{Convergence tests for the Cook's membrane for Maxwell (red square data on blue dashed curve) and Newtonian (red circle data on blue solid curve) models. We display the relative error on the computed $u_1$, as measured at the top-right corner of the deformed Cook's membrane, at time $t^{\star}=1$, versus the number of elements, $N_e=8,22,80,336$, both in logarithmic scale.}\label{fig:Convergence}
\end{figure}

\section[Results and Discussion]{Results and Discussion}\label{Sec4}
\subsection[Convergence Tests]{Convergence Tests}\label{Sec4-1}

We present our numerical results in absence of gravity, and in terms of the surface coordinate system, for which the surface vector displacement and applied loads only have two in-plane nontrivial components, and therefore we omit the null third component to avoid cumbersome notation. To validate our formulation and implementation, we have performed several convergence tests. One of the typical convergence tests for membrane structures is the Cook's membrane \cite{Brink1996,Simo}. This is a free-boundary problem in which a membrane, shaped as in figure \ref{fig:Cook}, undergoes a load (equally distributed among all nodes) on its top boundary, while its bottom boundary is held fixed. We apply a horizontal load along the top boundary, given by $\mb{P} = (1,0)$~N, and the bottom boundary satisfies a homogeneous Dirichelet boundary condition on the displacement field, i.e.~$\bs{u}=(0,0)$~m. All other boundaries of the membrane are free to move, and satisfy no-flux and traction-free boundary conditions. In problems in which a load is applied and/or removed, we relate the time of the loading/unloading phases to the characteristic time of the response of the material. For the case of Maxwell liquids we scale the time interval of the loading/unloading phases by the normalized time $t^{\star} = t / \tau$. On the other hand, for Newtonian liquids, we use $t^{\star} = t / t_c$, where $t_c = 1$~s is a characteristic time scale for viscous fluids. For the convergence test of both the Newtonian and Maxwell models, we apply a load with an amplitude $\mathcal{A}$, linearly varying in time, with $\mathcal{A}=1$ at $t^{\star}=0$ and $\mathcal{A}=0$ at $t^{\star}=1$. For the numerical investigations that follow, we consider membranes of viscosity coefficient $\eta = 10$~Pa~s, density $\rho = 10^3$~kg/m$^3$, and relaxation time $\tau =1$~s for the Maxwellian (for Maxwell model) membrane, unless specified differently. In figure \ref{fig:CookLastStep}, we show the deformed Maxwellian membrane at time $t^{\star} = 1$, discretized by an unstructured mesh composed of $336$ triangular elements. We display the contour plots of the vector displacement field, for which warmer shades indicate higher values. In figure \subref*{fig:CookLastStepU1}, we show the contour plot of the first component of the vector displacement field, $u_1$, that ranges between its minimum value, ${u_1}_{min} \sim0$~m (blue), on the bottom boundary, and its maximum value, ${u_1}_{max}=4.648 \times 10^{-3}$~m, (red) on the top-right corner of the membrane. In figure \subref*{fig:CookLastStepU2}, we display the contour plot of the second component of the vector displacement field, $u_2$, that ranges between its minimum value, ${u_2}_{min}=-3.275 \times 10^{-3}$~m (blue), on the top-right corner of the membrane, and its maximum value, ${u_2}_{max}=4.385\times10^{-4}$~m (red), on the left boundary.

By performing several numerical experiments, with a fixed time step, $\Delta t = 10^{-4}$~s, and refined unstructured meshes, we can have a quantitative analysis of the convergence of our numerical algorithms, equations (\ref{DiscreteNewtonianVector}) and (\ref{DiscreteMaxwellVector}), and show that our results converge under mesh refinement. Since the analytical solution of the particular free-boundary problem depicted in figure \ref{fig:Cook} is not known, it is a common practice to use the displacement field components, measured at one of the tips of the membrane, for convergence tests, as also described in the literature (see for instance, \cite{Brink1996,Simo}). In figure \ref{fig:Convergence}, we show our numerical results of the relative error on the computed $u_1$, as measured at the top-right corner of the deformed Cook's membrane, at time $t^{\star}=1$, versus the number of elements, $N_e=2,8,22,80,336$, both in logarithmic scale. For the computation of the relative error, we have considered as approximation of the actual solution, the results obtained with an unstructured mesh composed of $N_e=1346$ elements. We represent with red squares on a blue dashed curve the data for the Maxwellian membrane, and with red circles on a blue solid curve the Newtonian one. We can see that the results of our implementation of both constitutive models converge, with increasing number of elements.

\begin{figure}
\centering
\resizebox{0.4\textwidth}{!}{
\begin{tikzpicture}
\draw (0,0)--(3,0)node[anchor=south] {};
\draw (0,0)--(0,3)node[anchor=south] {};
\draw (0,3)--(3,3)node[anchor=south] {};
\draw (3,3)--(3,0)node[anchor=south] {};
\draw (-0.1,1.5)node[anchor=east]{$u_1=0$};
\draw (1.5,-0.1)node[anchor=north]{$u_2=0$};
\draw[->] (1.5,3)--(1.5,4) node[anchor=east] {};
\end{tikzpicture}%
}
\caption{Schematic of the pressure validation test.}\label{fig:setupTensionTest}
\end{figure}
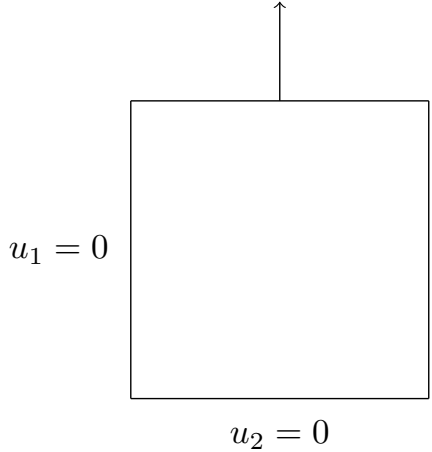

\begin{figure}[t]
\centering
\captionsetup{type=figure}
\subfloat[]{\includegraphics[scale=0.35,valign=t,trim=0.05in 0in 0.3in 0in,clip=true]{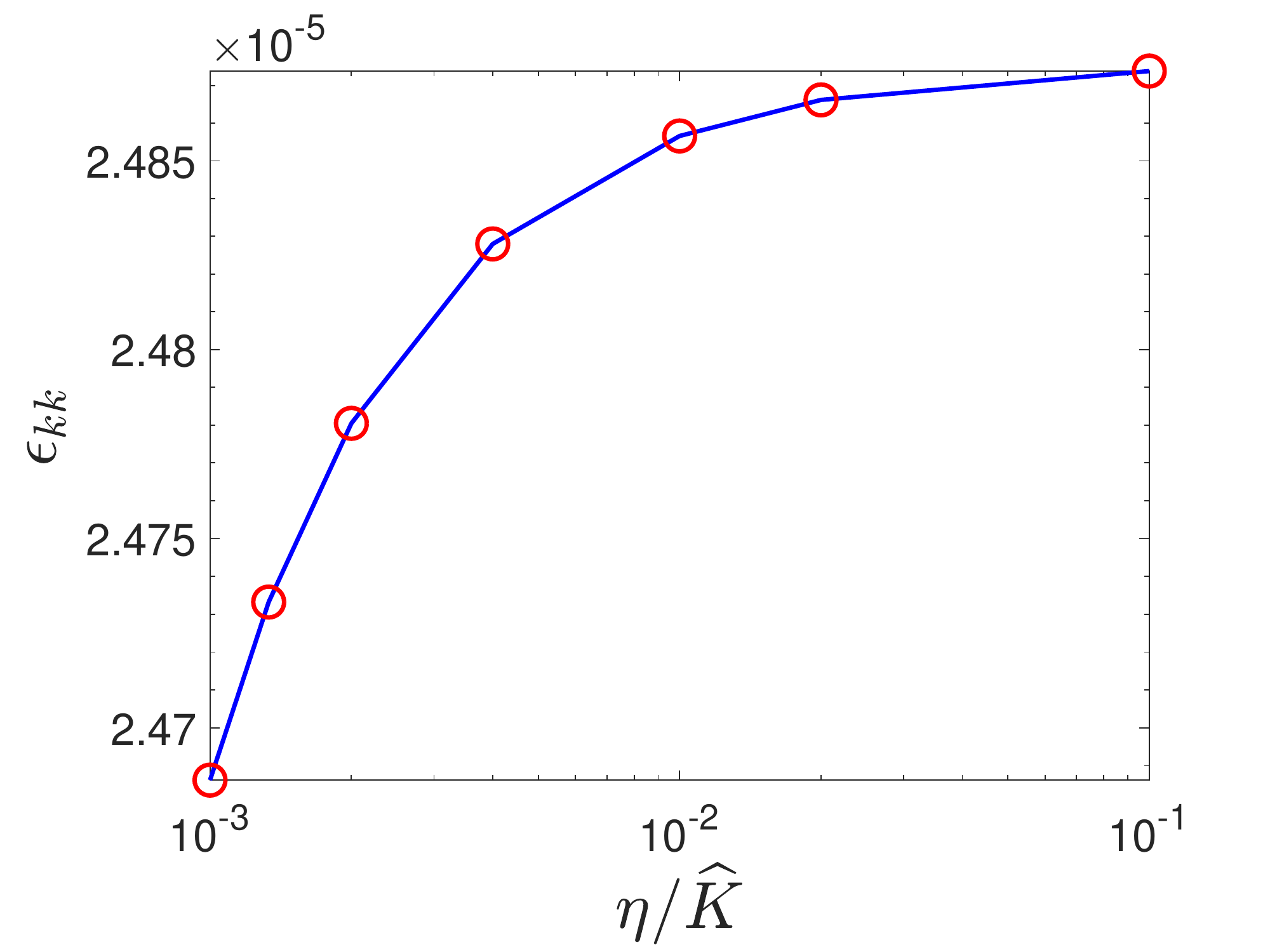}\label{fig:trE}}
\subfloat[]{\includegraphics[scale=0.35,valign=t,trim=0.05in 0in 0.3in 0in,clip=true]{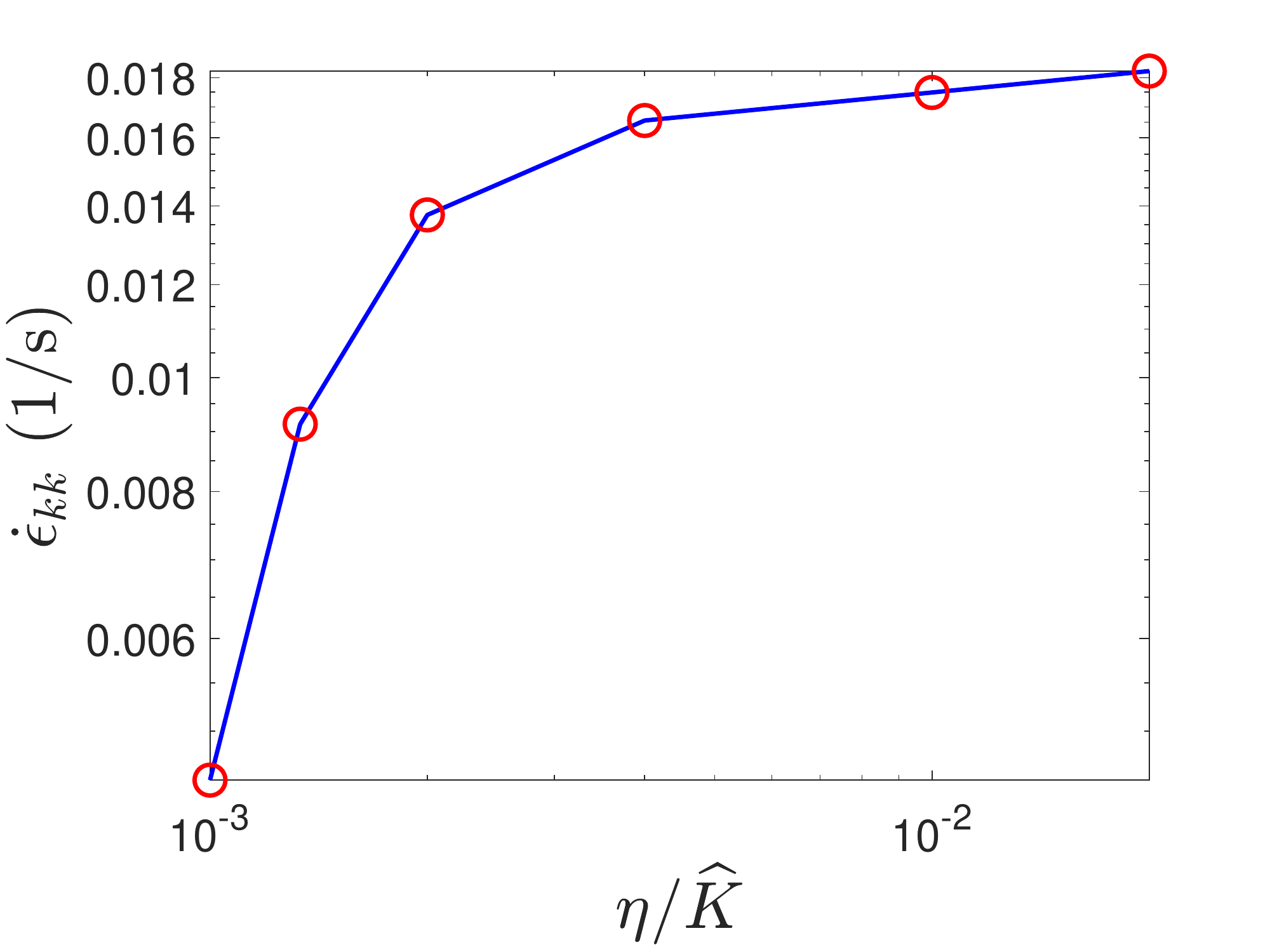}\label{fig:trEDot}}
\caption{In \protect\subref{fig:trE}, the computed volumetric strain, $\epsilon_{kk}$, and in \protect\subref{fig:trEDot}, the corresponding volumetric strain rate, $\dot{\epsilon}_{kk}$, versus the dimensionless constant $\eta/\widehat{K}$, for different values $\epsilon_{kk}$, versus the dimensionless constant $\eta/\widehat{K}$, for different values of $\widehat{K}=10^2,5\times10^2,10^3,5\times 10^3, 7.5 \times 10^3, 10^4$~Pa~s, keeping $\eta= 10$~Pa~s fixed, both in logarithmic scale.}\label{fig:trEOverKHat}
\end{figure}

A validation test for the pressure formulation and the near incompressibility condition is given by a tension experiment. For this test, the liquid membrane is deposited on a plane and surrounded by rigid plates, forming a square bounding box. As depicted in figure \ref{fig:setupTensionTest}, on the left boundary, the plate is allowed to move vertically, by imposing a zero boundary condition for $u_1$, while on the bottom boundary, the plate is allowed to move horizontally, by imposing a zero boundary condition for $u_2$. The right boundary satisfies a no-flux and traction-free boundary condition. The top boundary satisfies a no-flux boundary condition and it is displaced linearly in time by $\bs{u}=(0,0.005)$~m, such that for $t^{\star} = 0$, the corresponding amplitude is $\mathcal{A} =0$, and for $t^{\star}=1$, the corresponding amplitude is $\mathcal{A}=1$. Hence, by studying the dimensionless parameter related to the pressure, $\eta/\widehat{K}$, we can quantitatively verify that the pressure formulation leads, in the limit, to incompressibility. In figure \ref{fig:trEOverKHat}, we show the results of our simulations for the Maxwellian film. We measure the computed volumetric strain, ${\epsilon}_{kk}$ (shown in figure \subref*{fig:trE}), and the corresponding volumetric strain rate, $\dot{\epsilon}_{kk}$ (shown in figure \subref*{fig:trEDot}), for different values of $\widehat{K}=10^2,5\times10^2,10^3,5\times 10^3, 7.5 \times 10^3, 10^4$~Pa~s, keeping $\eta= 10$~Pa~s fixed, both in logarithmic scale. We can see that the data corresponding to the small values of the dimensionless ratio $\eta/\widehat{K}$, have both smaller volume change, $\epsilon_{kk}$, and respective rate of change, $\dot{\epsilon}_{kk}$. We have found that the optimal range for the near incompressibility condition is $\eta/\widehat{K} \in [10^{-3},10^{-1}]$. For values outside of this range the penalty method leads to stringent constraints on the time step \cite{Shen,LuEtAl}, or larger compressibility of the material.

\begin{figure}[t]
\centering

\centering
\resizebox{.35\textwidth}{!}{
\begin{tikzpicture}
\node (A) at (0,0){};
\node (A') at (-0.1,0.05){};
\node (A'') at (-0.2,0.1){};
\node (B) at (1,0.3){};
\node (C) at (1,1.3){};
\node (C') at (0.9,1.35){};
\node (C'') at (0.8,1.4){};
\node (D) at (0,1){};
\node (D') at (-0.1,1.05){};
\node (D'') at (-0.2,1.1){};
\draw[thin] (0,0) coordinate(A) -- (1,0.3) coordinate(B) -- (1,1.3) coordinate(C) -- (0,1) coordinate(D) -- (0,0)coordinate(A);
\path[name path=border-mid1,fill=cyan!30] (-0.1,0.05) -- (-0.005,0.08)  -- (-0.005,1.005) -- (0.9,1.275) -- (.9,1.35) -- (-0.1,1.05) -- (-0.1,.05);
\draw[thin] (-0.2,0.1) coordinate (A'') -- (-0.2,1.1) coordinate (D'') -- (0.8,1.4) coordinate (C'');
\draw[thin] (-0.2,0.1) coordinate (A'') -- (-0.1,0.13);
\draw[thin] (0.8,1.4) coordinate (C'') -- (.8,1.33);

\draw[->] (0.0,1.31)--(0.7,1.52) node[anchor=east] {};
\end{tikzpicture}%
}
\caption{Schematic of a sheared membrane between parallel plates. Both plates are sheared on the top, held fixed at the bottom, and no-flux and traction-free boundary conditions are applied on the lateral boundaries of the plates. Friction between the liquid and the plates is neglected.}\label{fig:ParallelPlates}%
\end{figure}
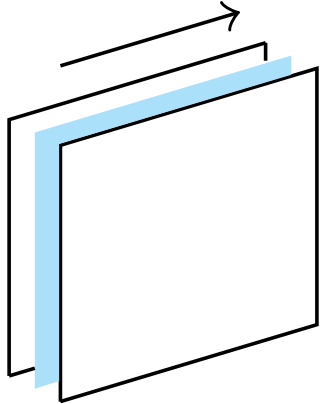

\subsection[Membrane deformation under shear flow]{Membrane deformation under shear flow}\label{Sec4-2}

\setlength\fboxsep{0 pt}
\pdfpxdimen=\dimexpr 1 in/300\relax

\begin{figure}[t]
{\includegraphics[width=4.2cm,valign=t,trim=0.7in 0.0in 2.8in 1in]{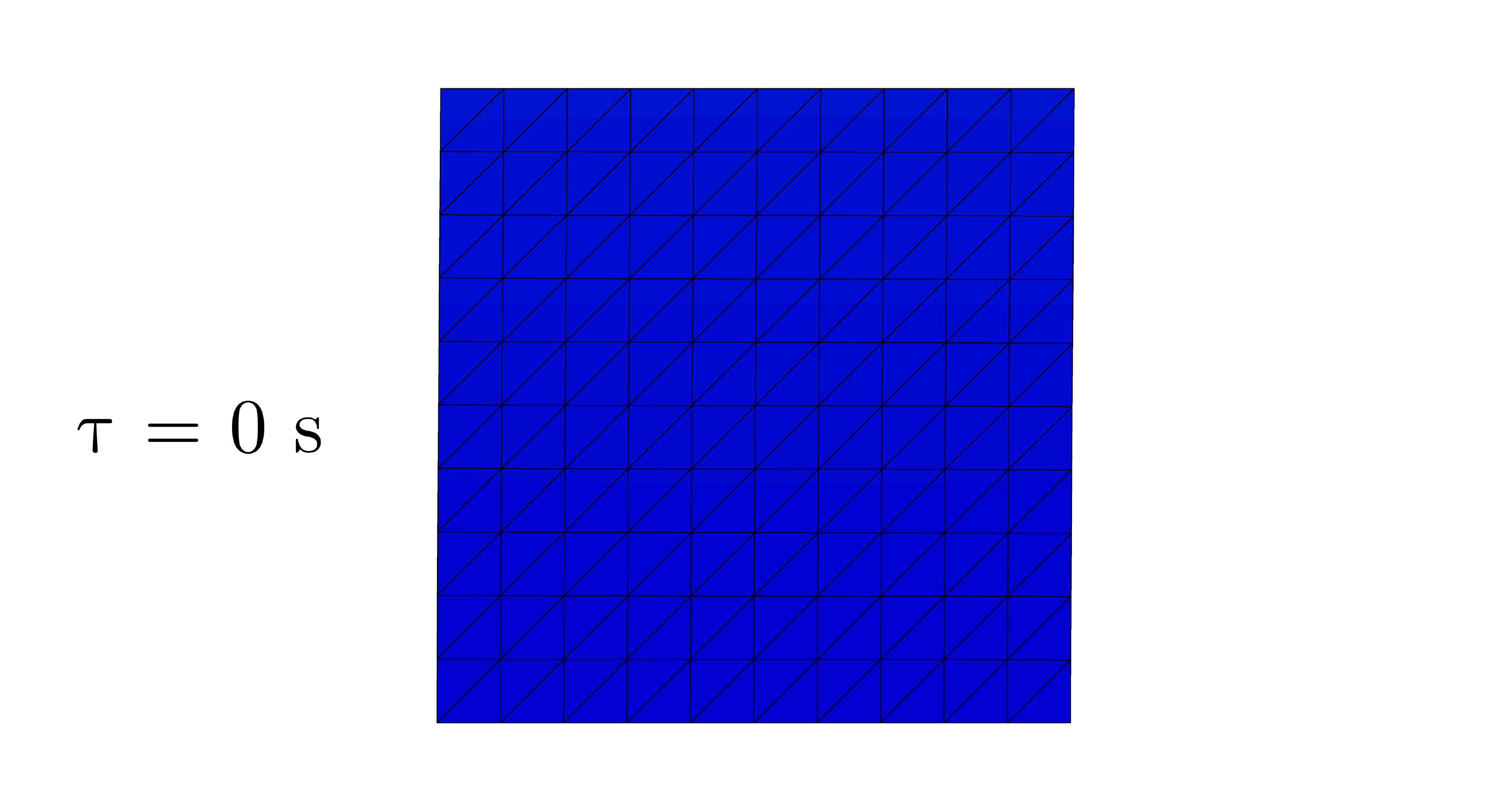}}
{\includegraphics[width=4.2cm,valign=t,trim=0.7in 0.0in 2.8in 1in]{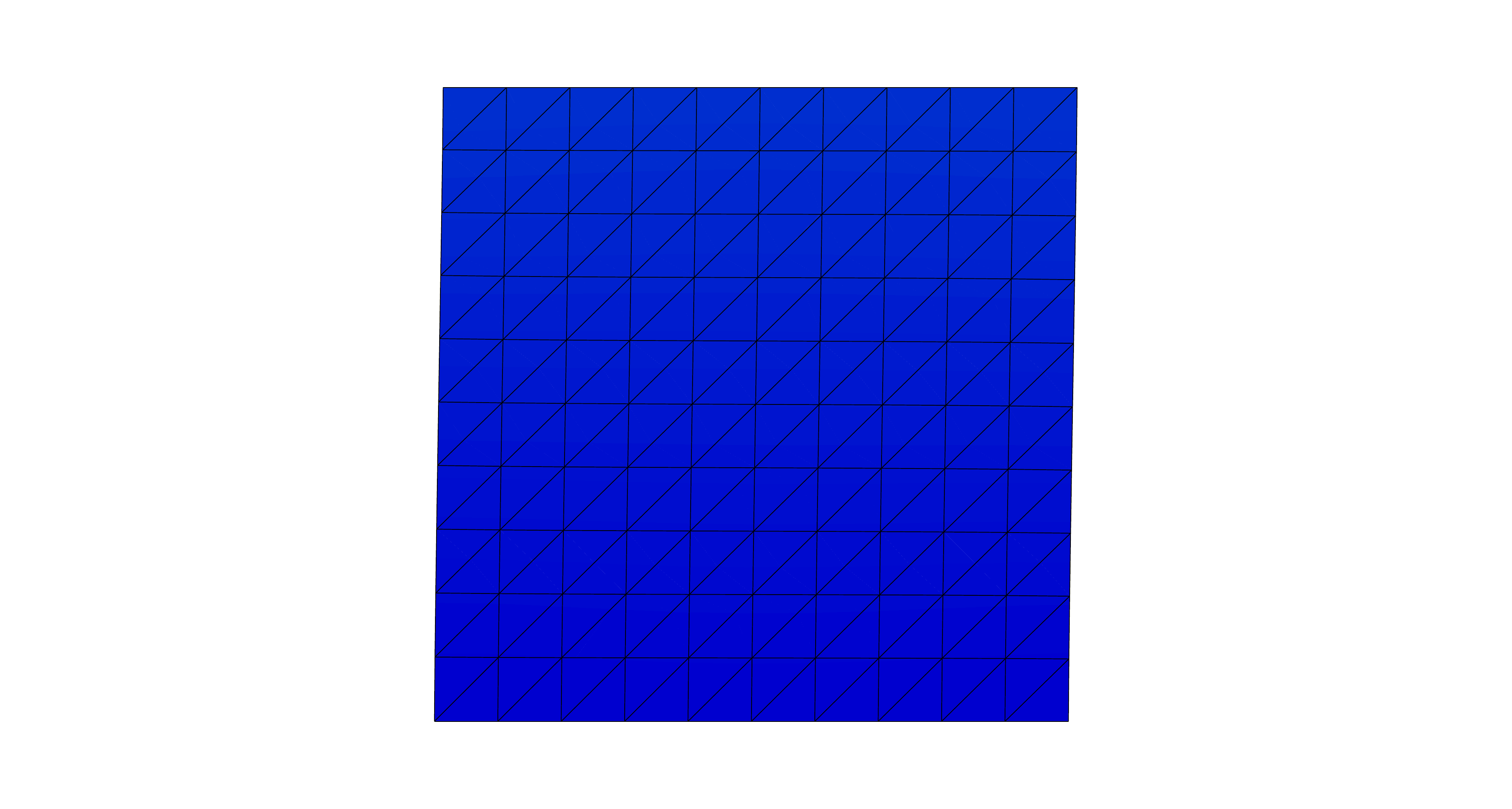}}
{\includegraphics[width=4.2cm,valign=t,trim=0.7in 0.0in 2.8in 1in]{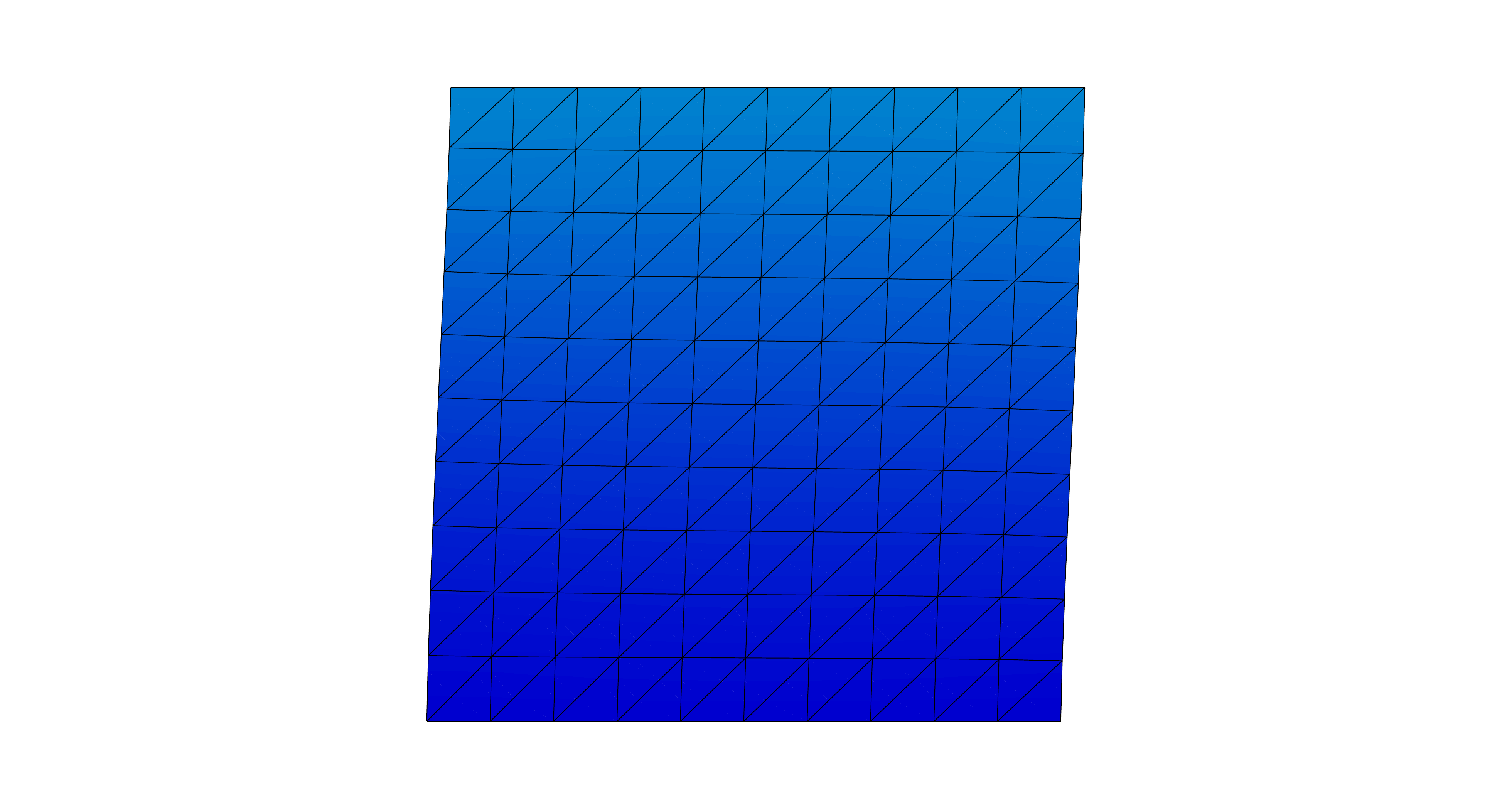}}\\
{\includegraphics[width=4.2cm,valign=t,trim=0.7in 0.0in 2.8in 1in]{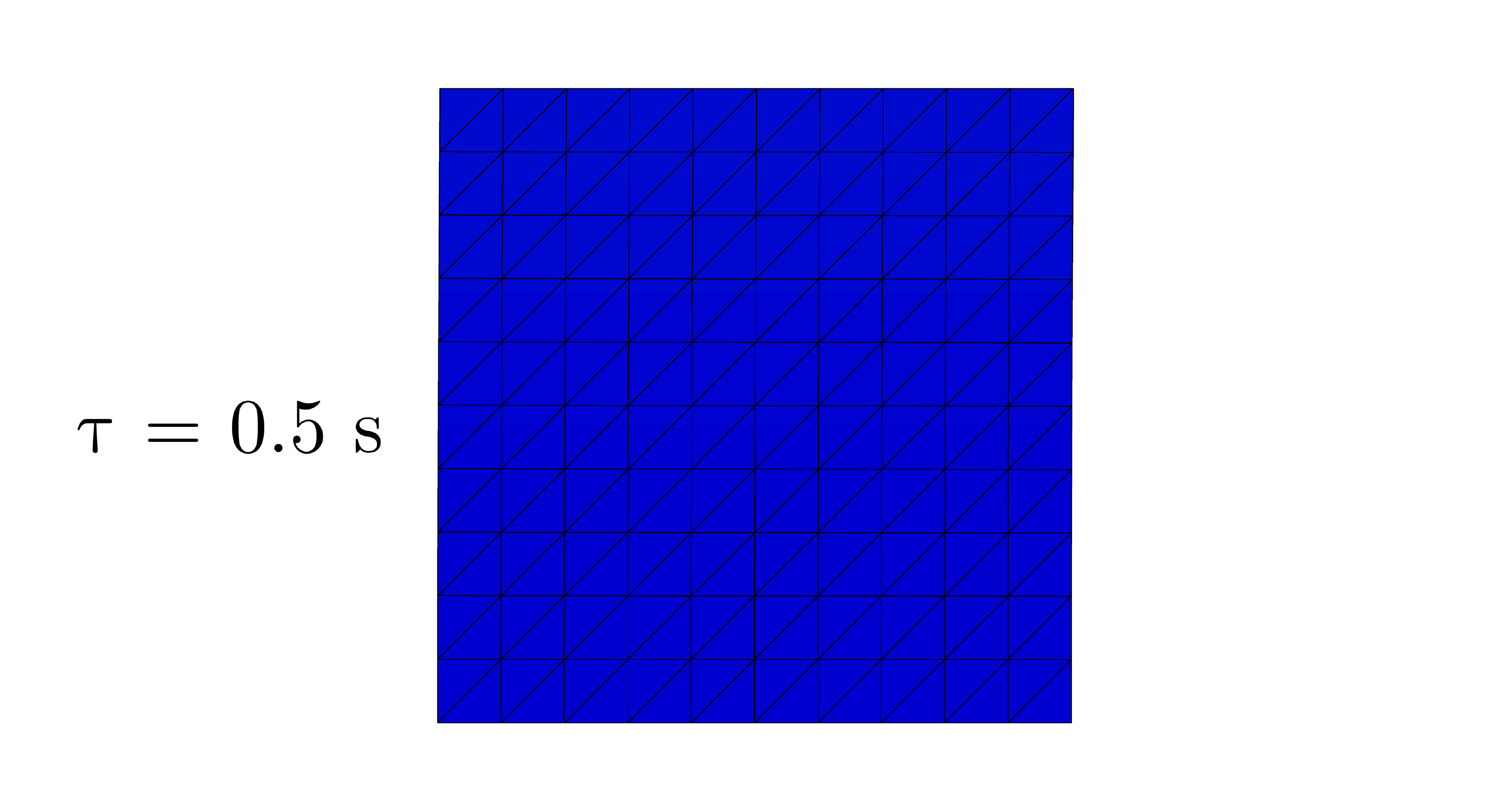}}
{\includegraphics[width=4.2cm,valign=t,trim=0.7in 0.0in 2.8in 1in]{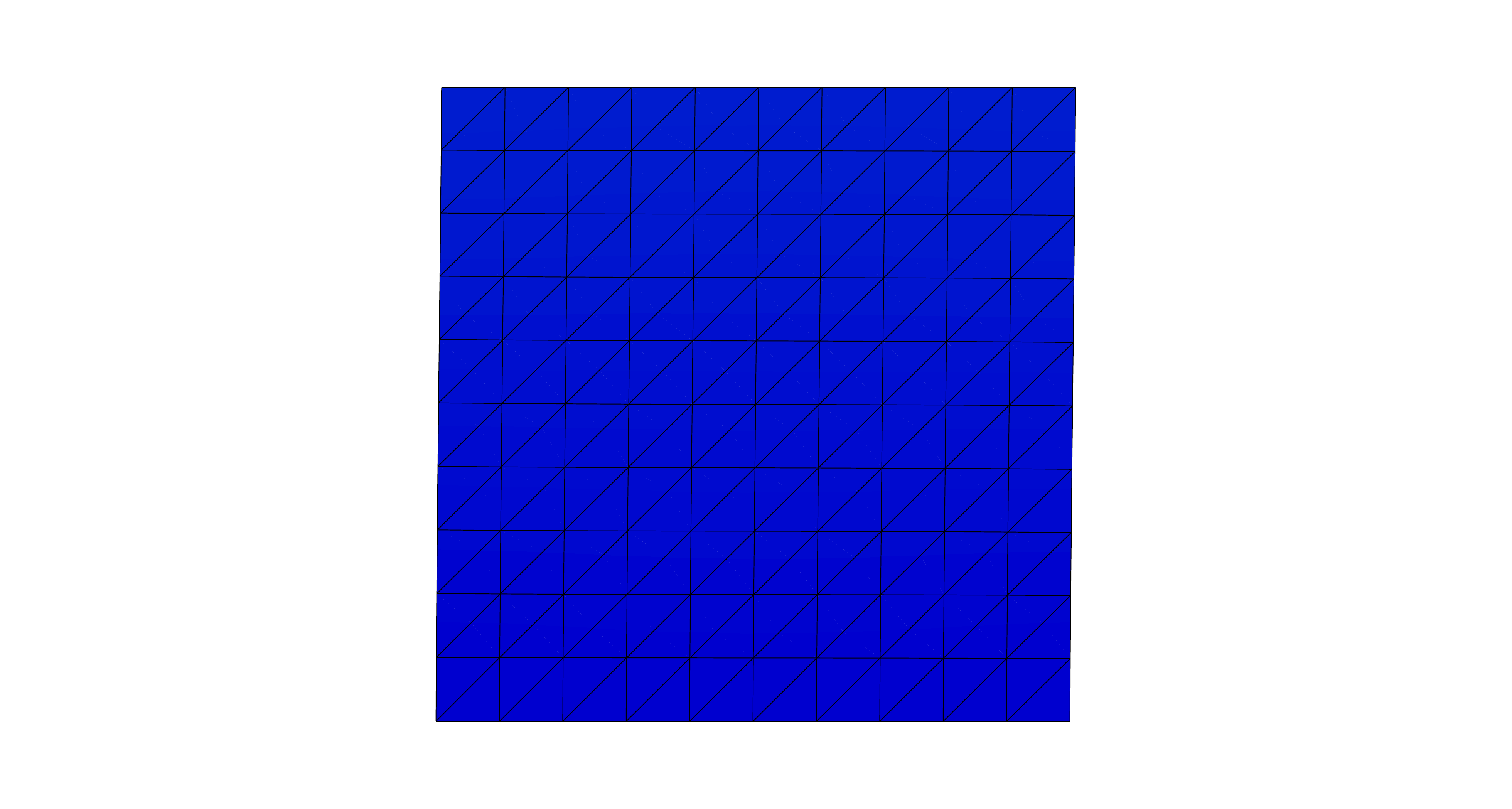}}
{\includegraphics[width=4.2cm,valign=t,trim=0.7in 0.0in 2.8in 1in]{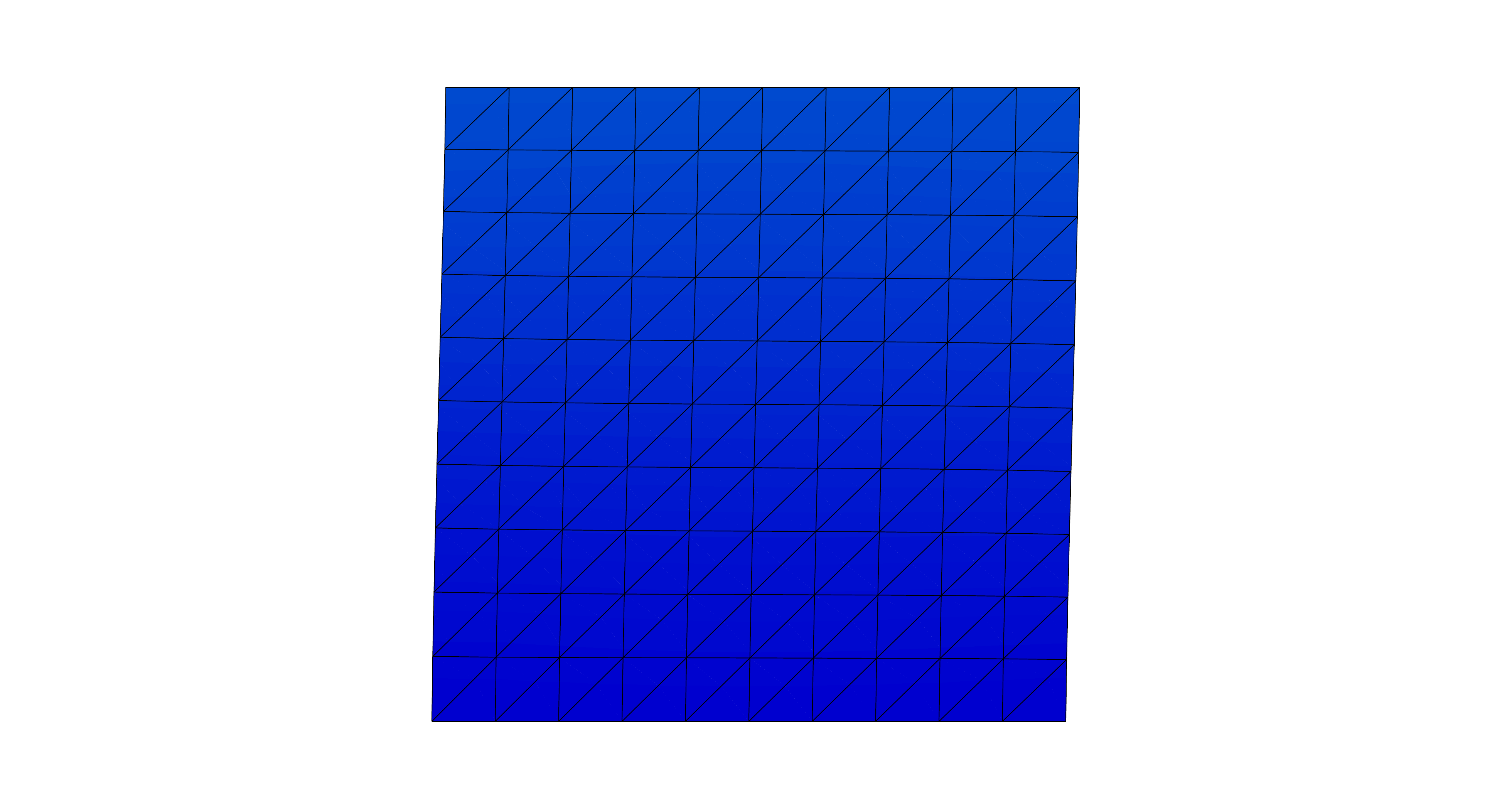}}\\
{\includegraphics[width=4.2cm,valign=t,trim=0.7in 0.0in 2.8in 1in]{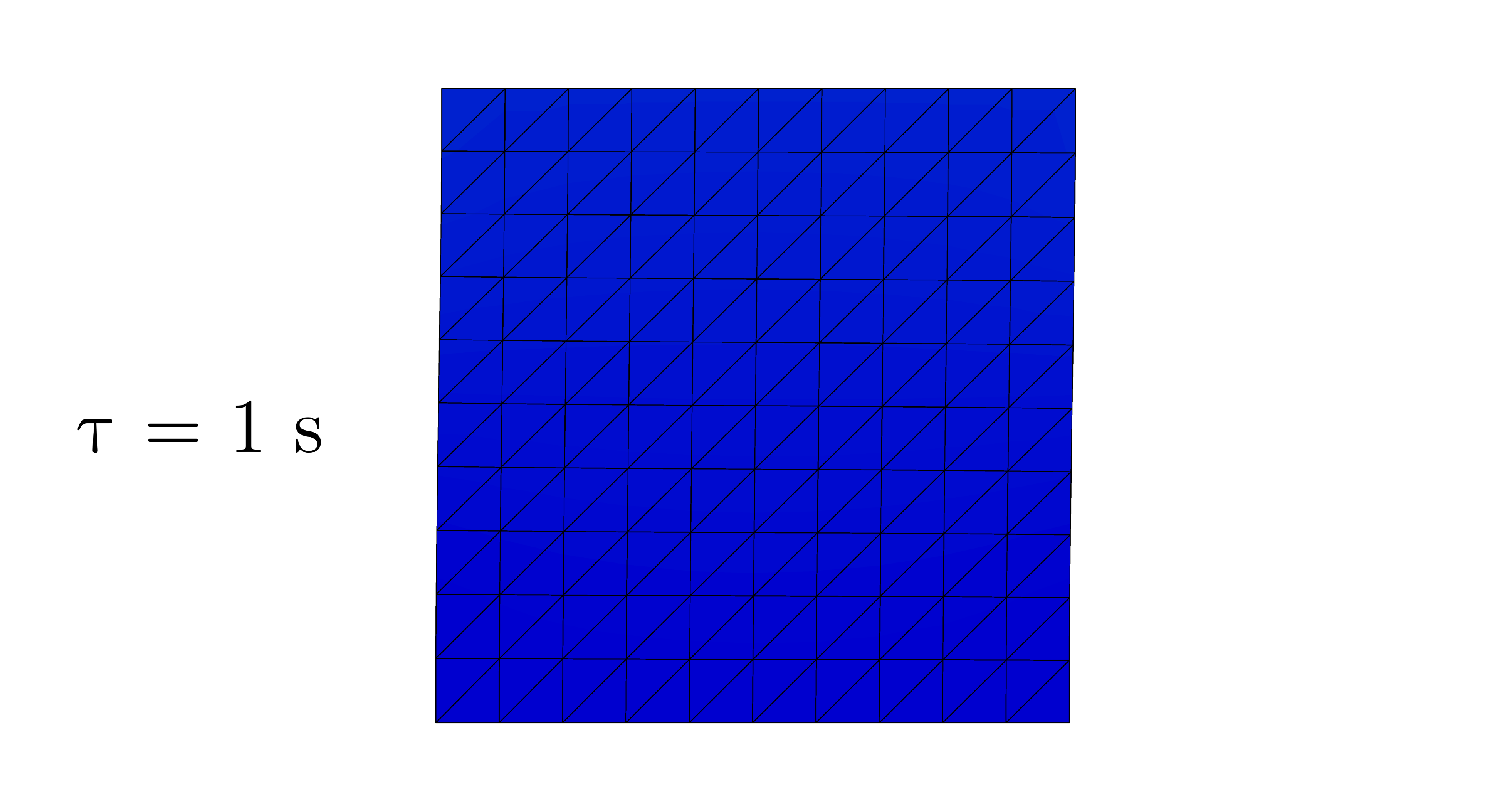}}
{\includegraphics[width=4.2cm,valign=t,trim=0.7in 0.0in 2.8in 1in]{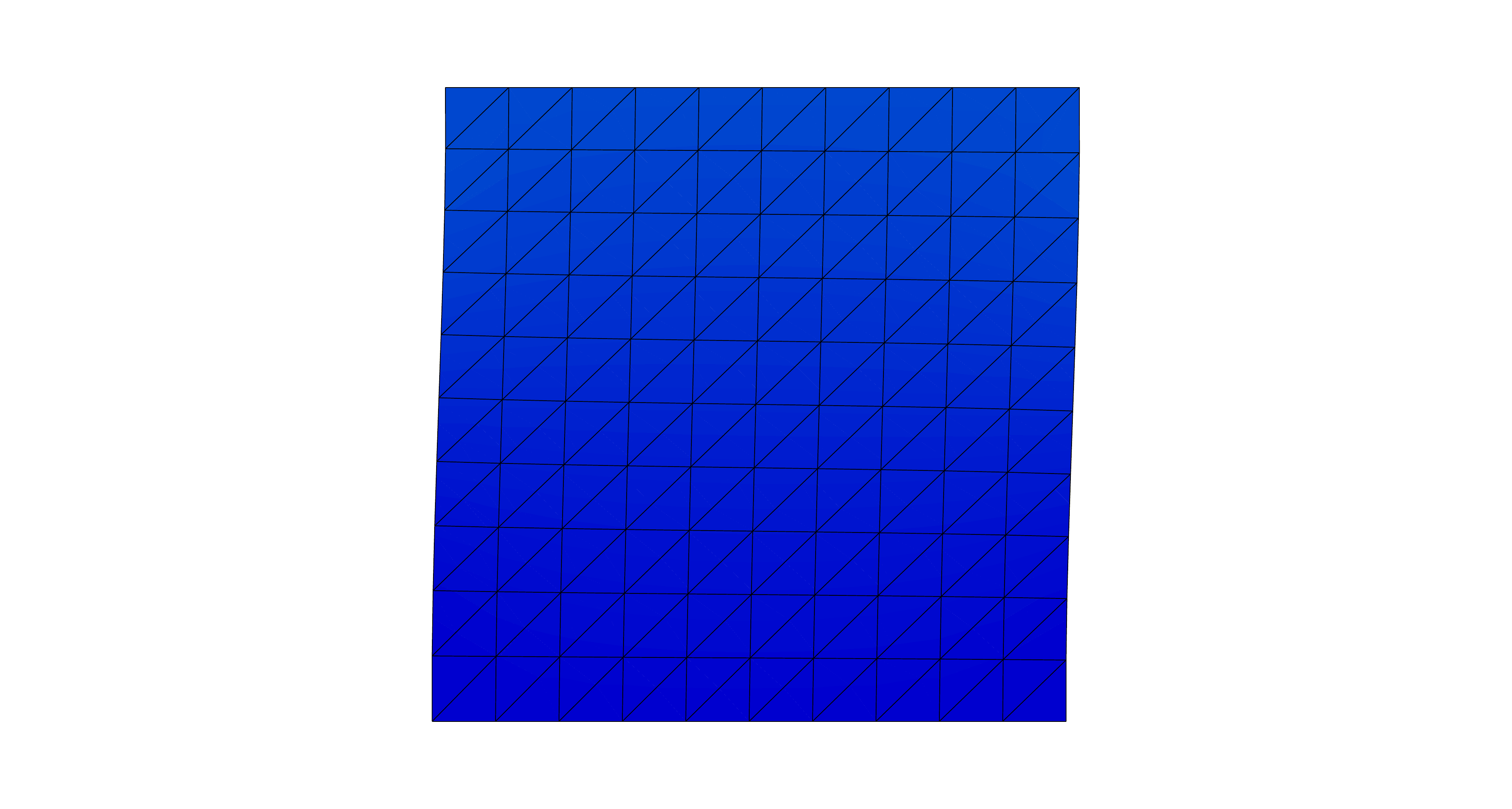}}
{\includegraphics[width=4.2cm,valign=t,trim=0.7in 0.0in 2.8in 1in]{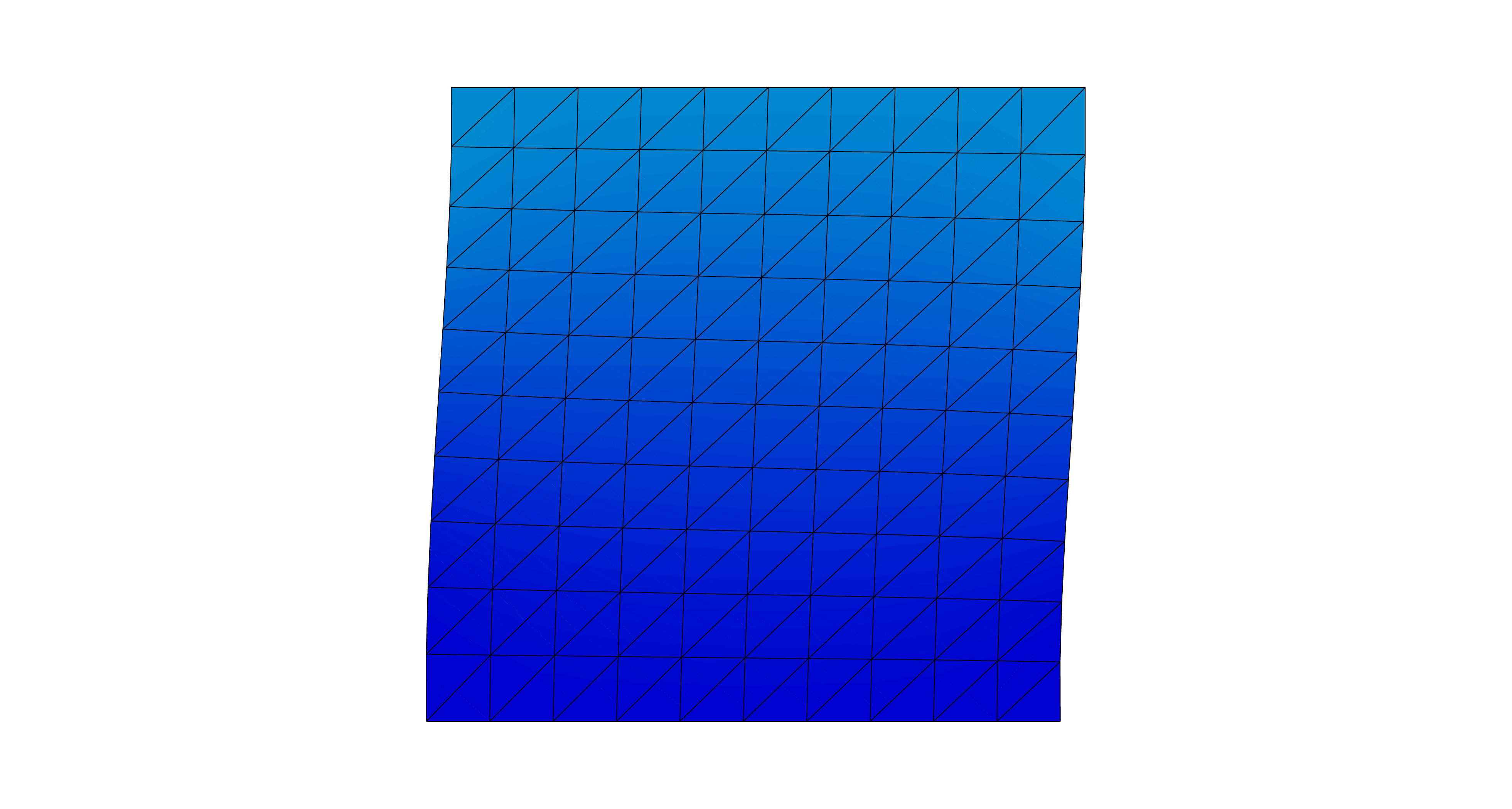}}\\
{\includegraphics[width=4.2cm,valign=t,trim=0.7in 0.0in 2.8in 1in]{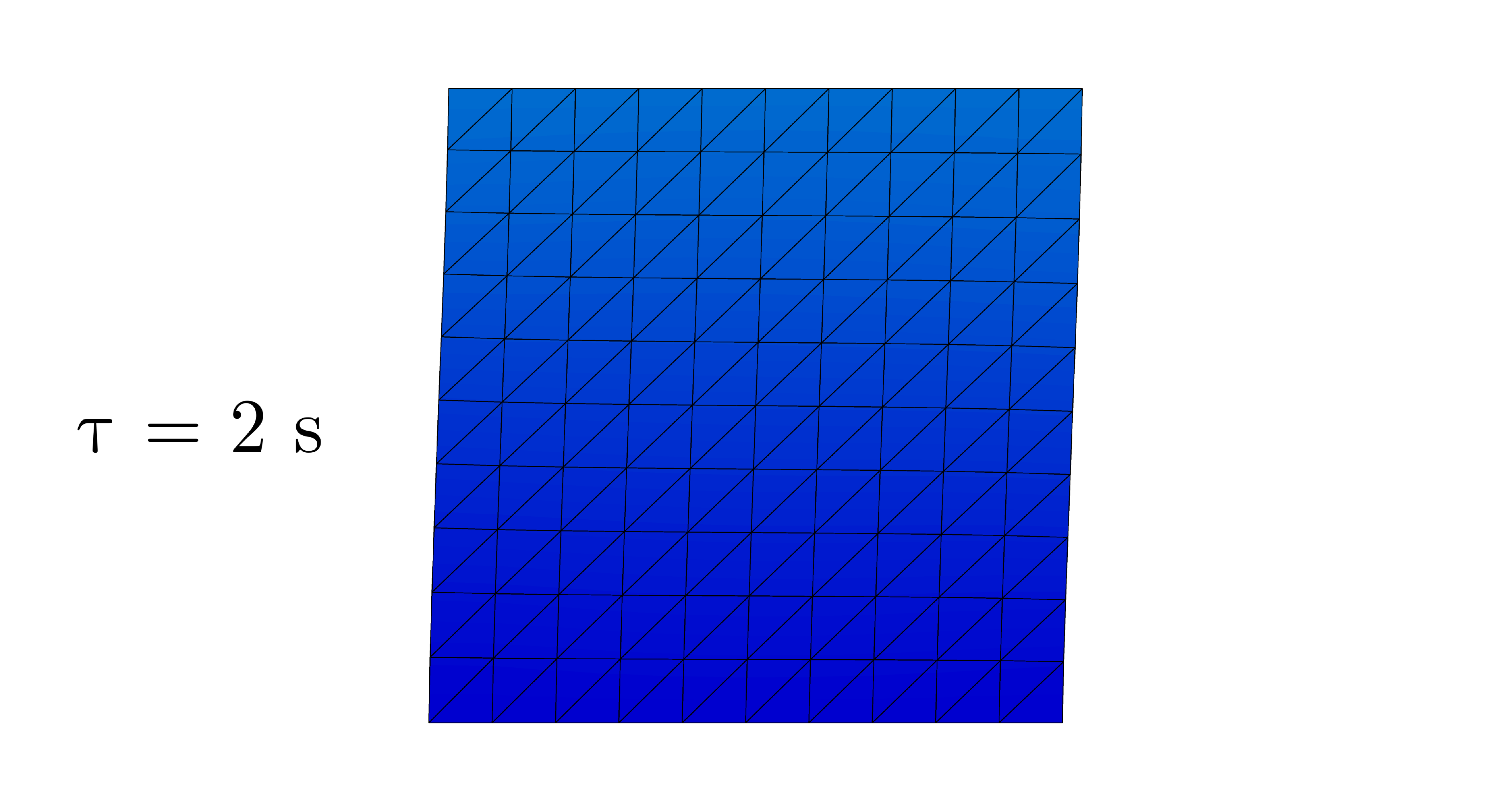}}
{\includegraphics[width=4.2cm,valign=t,trim=0.7in 0.0in 2.8in 1in]{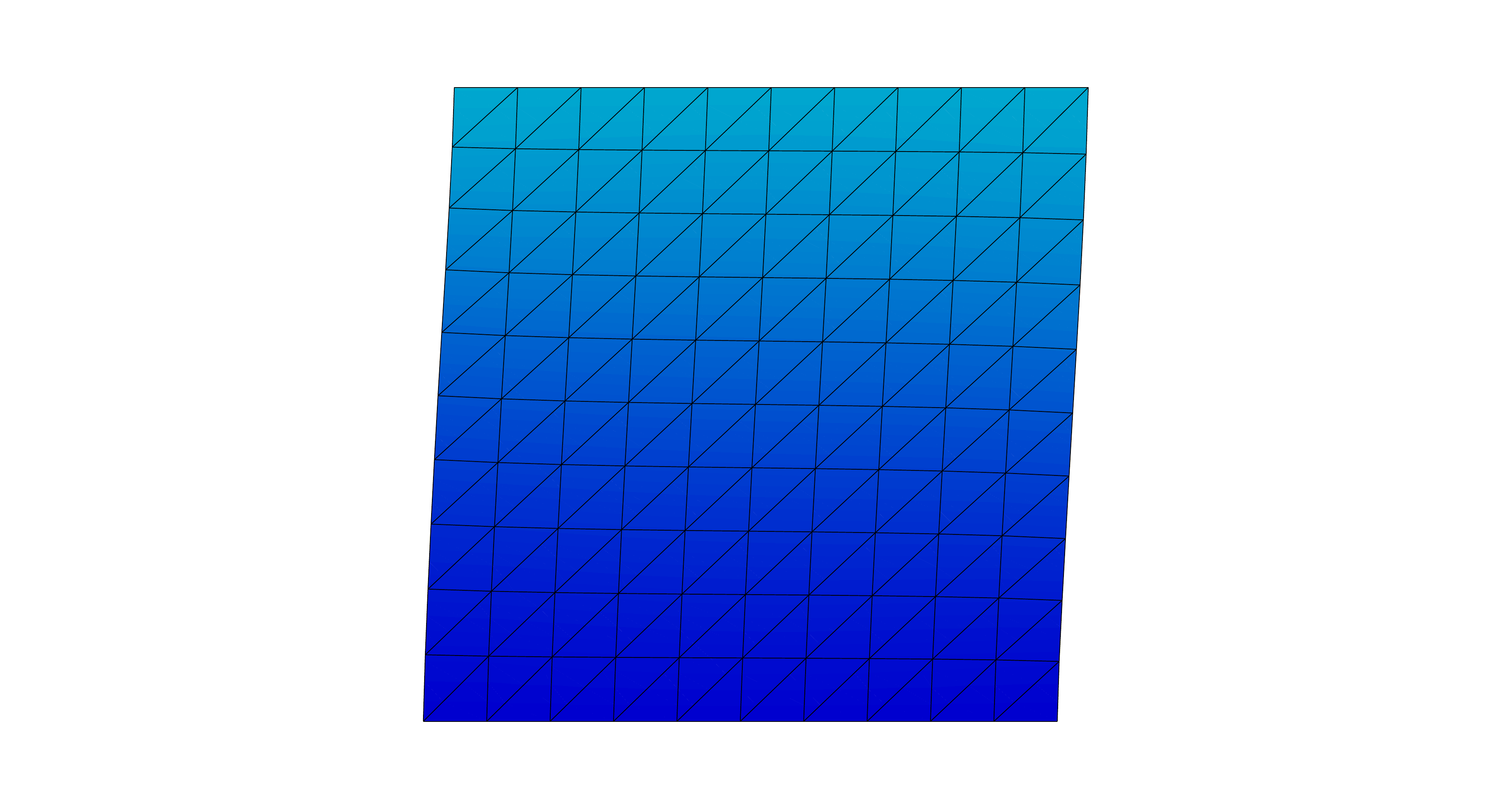}}
{\includegraphics[width=4.2cm,valign=t,trim=0.7in 0.0in 2.8in 1in]{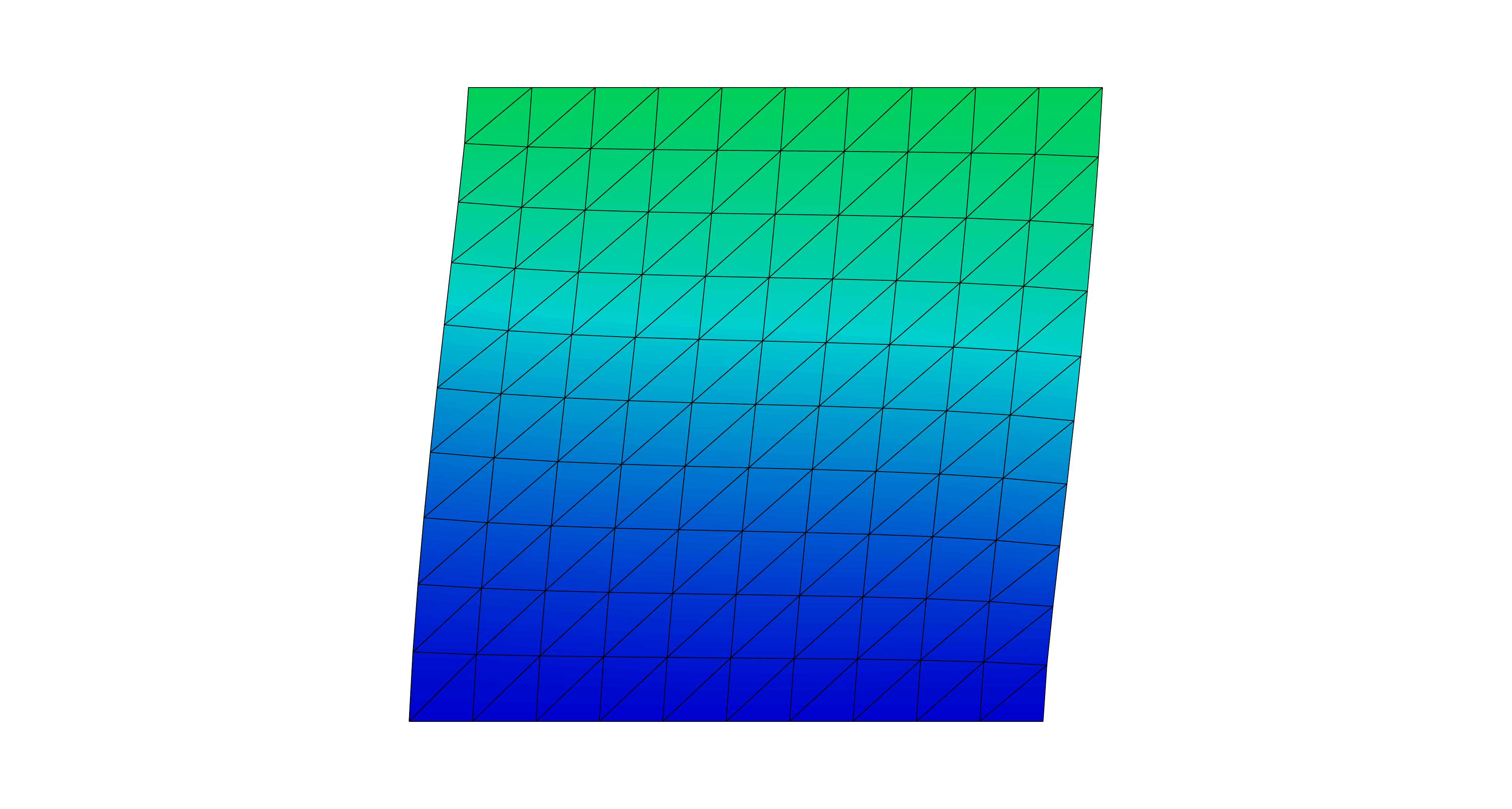}}\\
{\includegraphics[width=4.2cm,valign=t,trim=0.7in 0.1in 2.8in 1in]{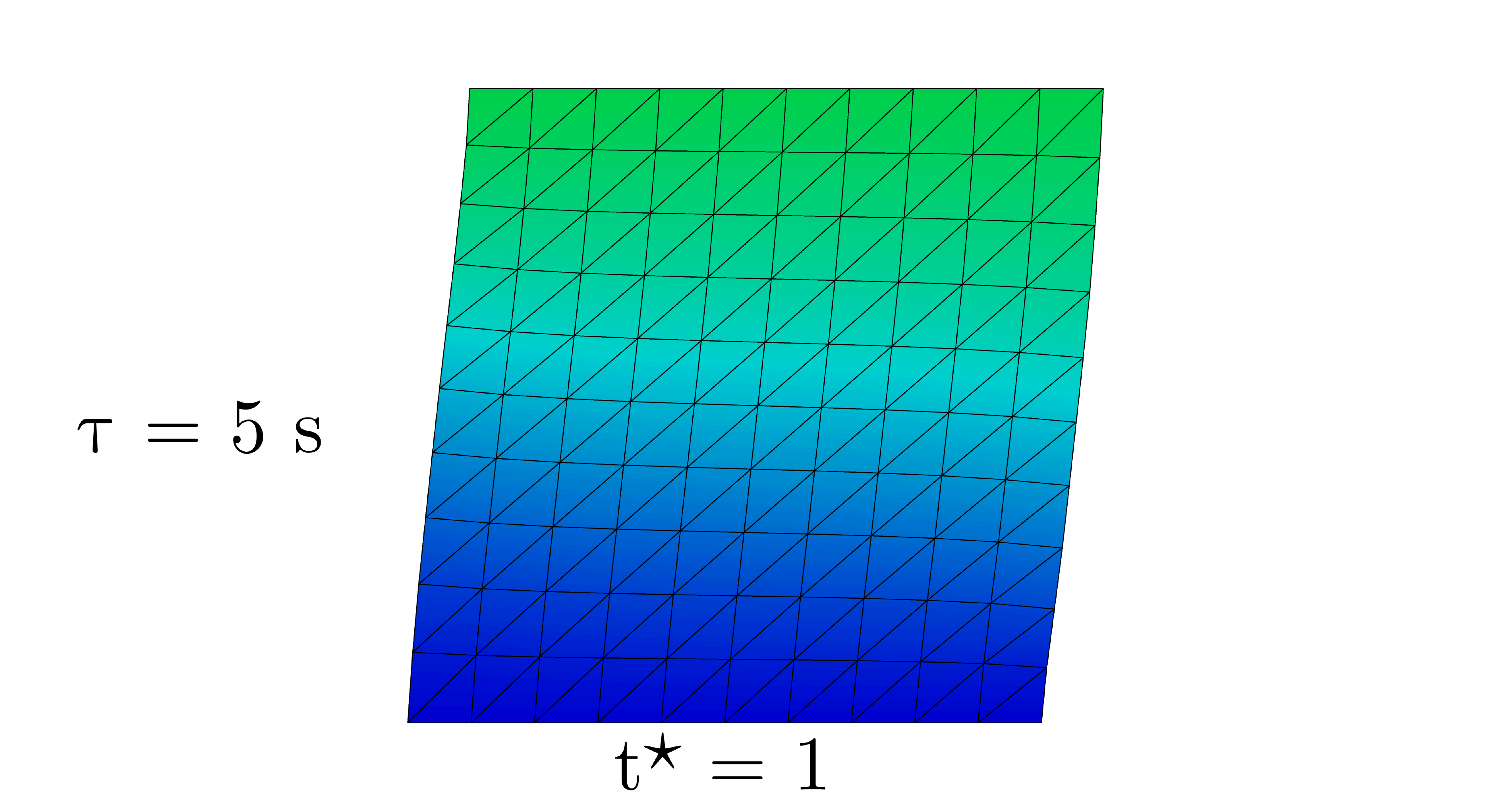}}
{\includegraphics[width=4.2cm,valign=t,trim=0.7in 0.1in 2.8in 1in]{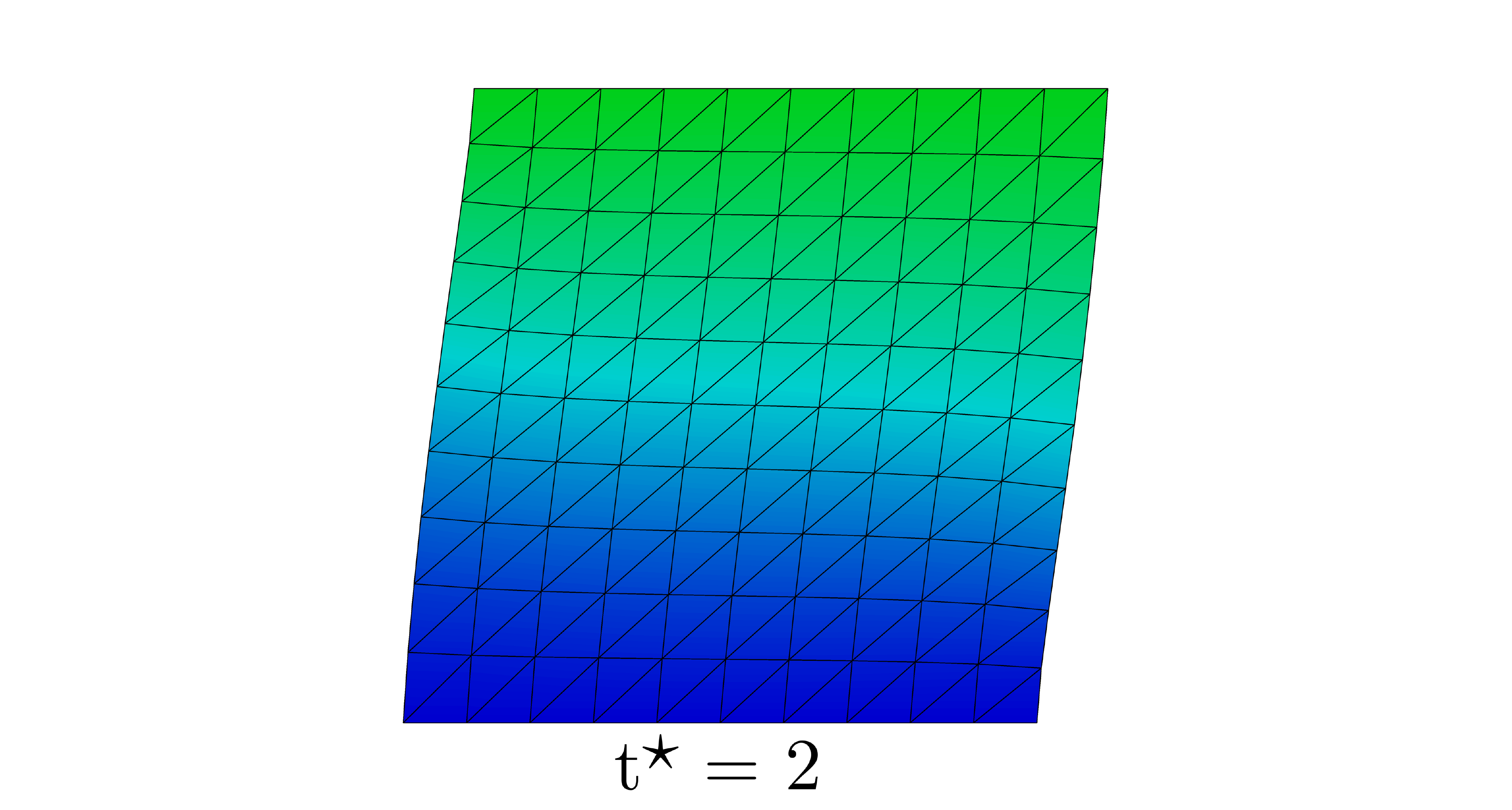}}
{\includegraphics[width=4.2cm,valign=t,trim=0.7in 0.1in 2.8in 1in]{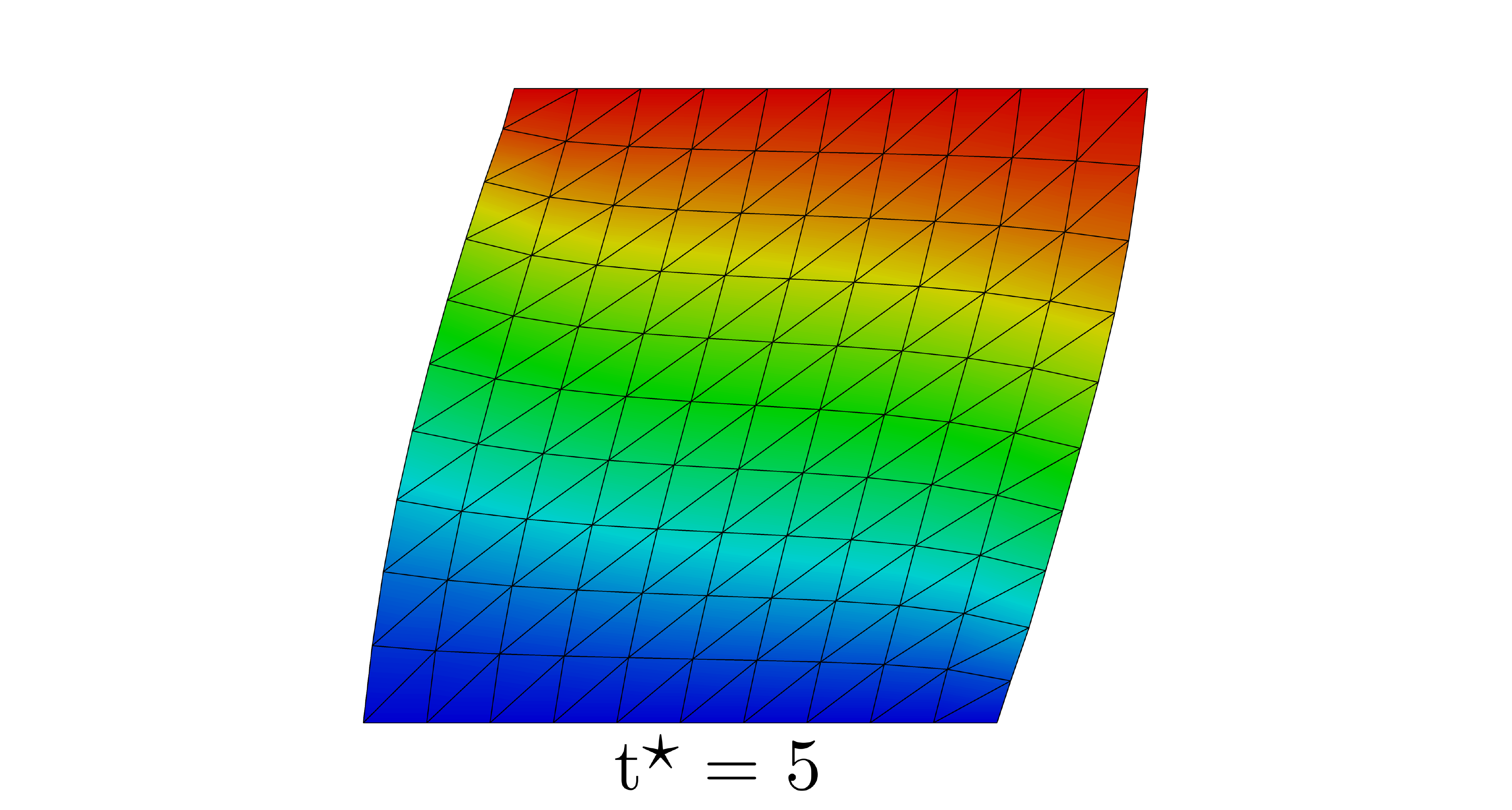}}
{\includegraphics[height=2.55cm,valign=t,trim=0cm 0cm 0.9cm 0.7cm]{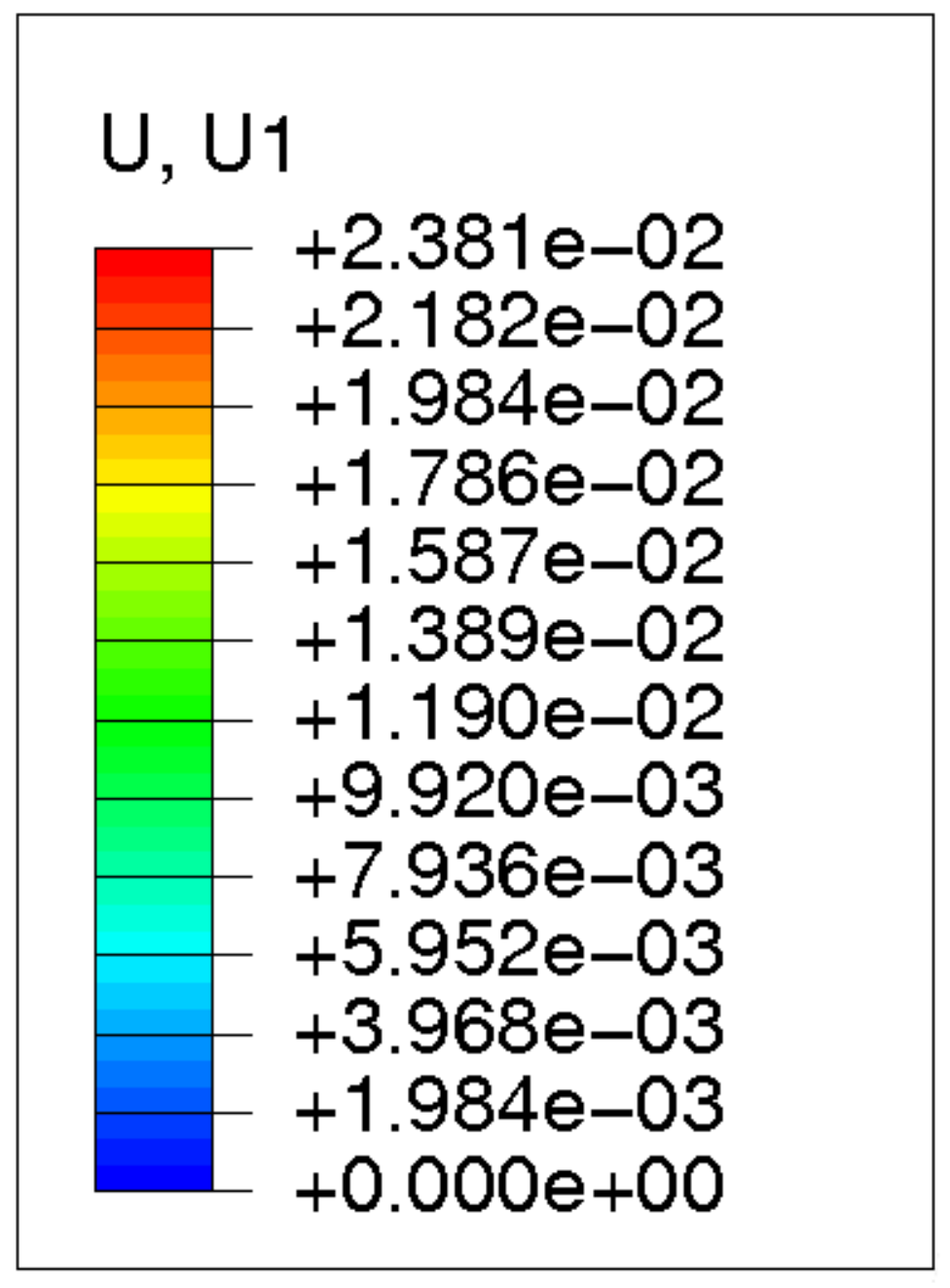}}
\caption{Comparison of the evolution of sheared membranes of Newtonian and Maxwellian fluids. The contour plots of $u_1$, are displayed at time $t^{\star}=1,2,5$ (from left to right), where the red color represents the maximum value attained at $t^{\star}=5$, ${u_1}_{max}=2.381 \times 10^{-2}$~m, and the blue color represents the minimum value, ${u_1}_{min}=0$~m. The viscosity coefficient for all membranes is $\eta = 20$~Pa~s, and the relaxation times are $\tau=0,0.5,1,2,5$~s, (from top to bottom).}\label{fig:SimpleShearComparison}
\end{figure}

The first application we consider is the simple shear flow \cite{Batchelor,Bird} of a thin liquid layer between parallel rigid plates. In this investigation, we do not include friction effects between the liquid layer and the rigid plates. The shear motion is obtained by holding fixed the bottom boundary of the plates, and by horizontally shearing the top boundary, by either imposing a horizontal displacement, or a horizontal force. In figure \ref{fig:ParallelPlates}, we show the schematic of the setup of this numerical experiment, where square membranes of length $L=10^{-1}$~m are used. For the first numerical experiment, a constant horizontal load $\mb{P}= (10^{-2}, 0)$~N has been linearly applied in time for $t^{\star} = 5$. The right and left boundaries satisfy a traction-free and no-flux boundary conditions, the bottom boundary is clamped, and the top is allowed to move only horizontally, by imposing the condition that $u_2=0$ on all nodes along the top boundary. In figure \ref{fig:SimpleShearComparison}, we show the final configuration of sheared membranes of Maxwell type, compared to a viscous one. The contour plots of $u_1$, are displayed at time $t^{\star}=1,2,5$ (from left to right), where the red color represents the maximum value attained at $t^{\star}=5$, ${u_1}_{max}=2.381 \times 10^{-2}$~m, and the blue color represents the minimum value, ${u_1}_{min}=0$~m. The viscosity coefficient for all membranes is $\eta = 20$~Pa~s, and the relaxation times are $\tau=0,0.5,1,2,5$~s, (from top to bottom). We observe that the liquid membrane of Maxwell type with the highest relaxation time has deformed the most, corresponding to a longer dimensional time of imposed load. Moreover, we notice how the Newtonian membrane (shown on the second row) is the one that displaces the least, compared to all other Maxwellian membranes, and therefore exhibits the darkest shades.

\begin{figure}[t]
\centering
\subfloat[]{\includegraphics[height=5cm]{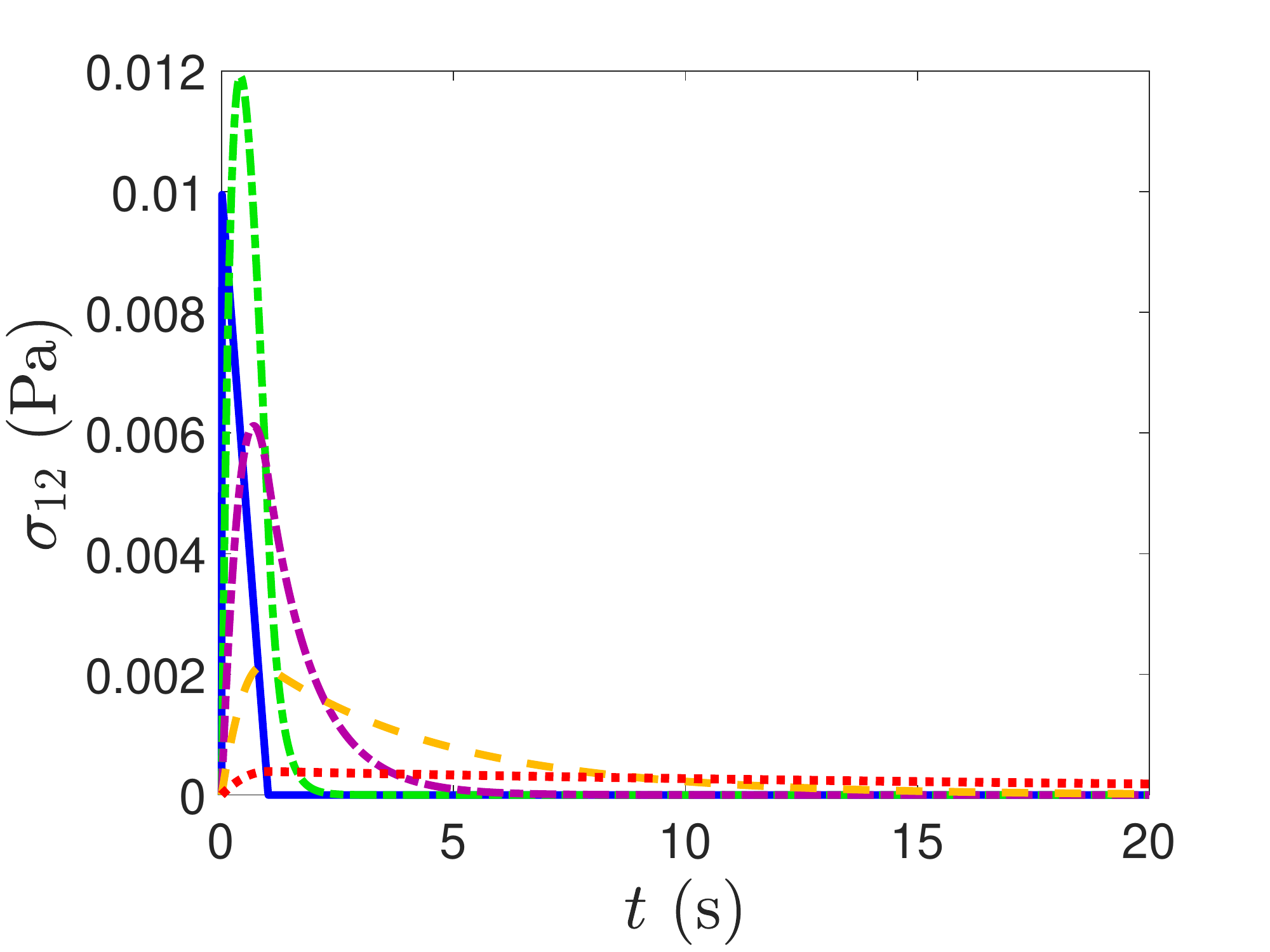}\label{fig:StressVsTime_a}}
\subfloat[]{\includegraphics[height=5cm]{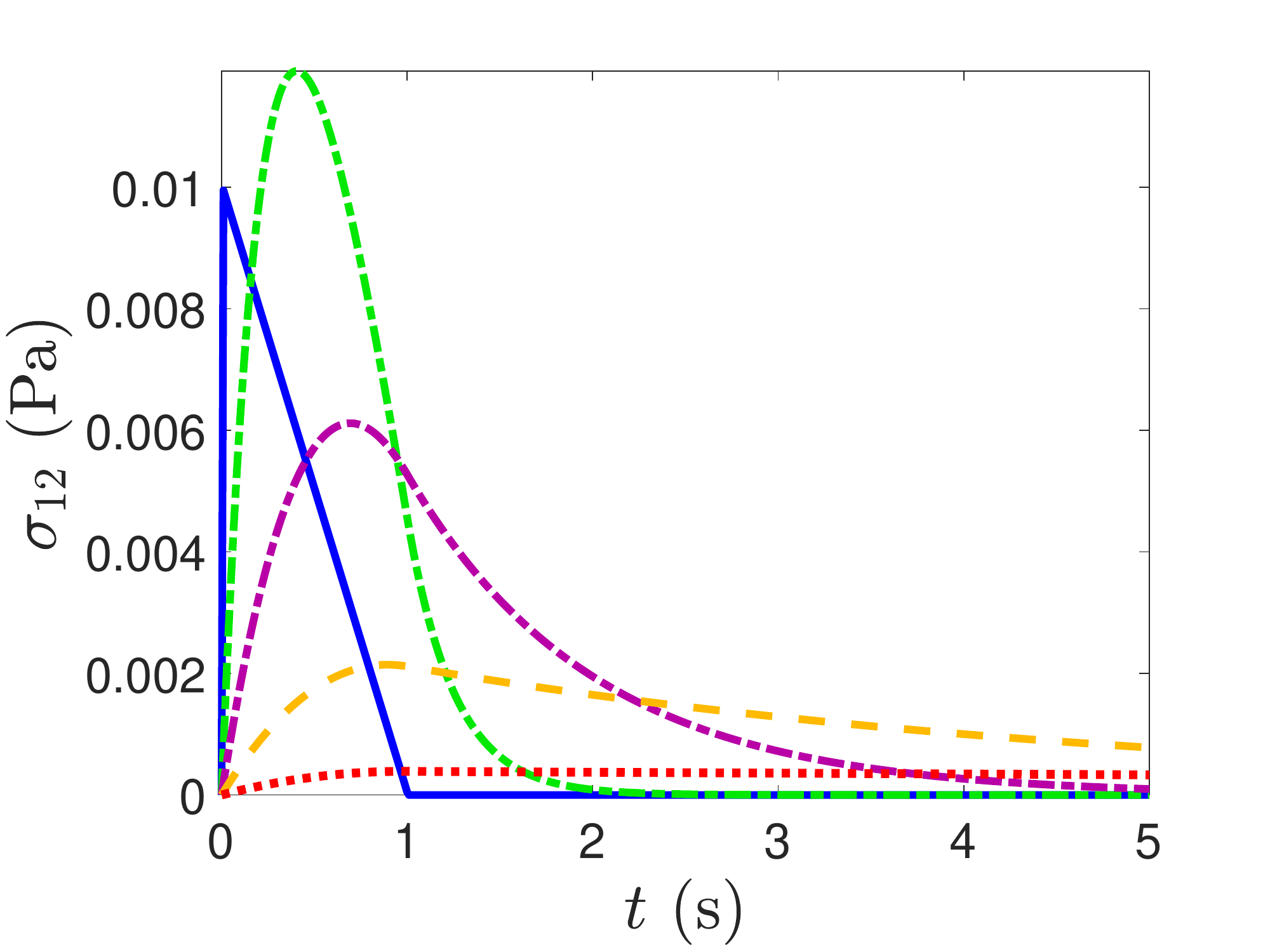}\label{fig:StressVsTime_b}}
\caption{\protect\subref{fig:StressVsTime_a} Evolution of the shear stress component, $\sigma_{12}$, of sheared membranes, for different values of the relaxation time $\tau = 0$~s (blue solid curve), $0.5$~s (green dashed curve), $1$~s (purple dash-dotted curve), $2$~s (yellow dashed curve), $5$~s (red dotted curve). \protect\subref{fig:StressVsTime_b} A magnification for $t\in[0,5]$~s.}\label{fig:StressVsTime}%
\end{figure}

Next, we investigate the effect of the relaxation time on both the stress and the displacement in the simple shear flow. We observe that the relaxation time, $\tau = \eta / G$, represents the ratio of the shear viscosity coefficient over the shear elastic modulus. Hence by keeping the viscosity fixed, and by increasing $\tau$, we increase the importance of viscosity relative to elasticity. In this test case, we displace the top plates by applying a velocity of $\textbf{v} = (10^{-4},0)$~m/s. This boundary condition is time-dependent, with the magnitude of the applied velocity linearly decreasing in time, with $\mathcal{A}=1$ at $t^{\star}=0$ and $\mathcal{A}=0$ at $t^{\star}=4$. In figure \ref{fig:StressVsTime}, we plot the time evolution (in \subref*{fig:StressVsTime_a}, for $t\in[0,20]$~s, and in \subref*{fig:StressVsTime_b}, a close-up for $t\in[0,5]$~s) of the shear stress component, $\sigma_{12}$, for the values of the relaxation time $\tau = 0$~s (blue solid curve), $0.5$~s (green dashed curve), $1$~s (purple dash-dotted curve), $2$~s (yellow dashed curve), $5$~s (red dotted curve), for a $2$-element test membrane, in which the stress is uniform and the same in both elements. In this figure we can see that the limiting case, for $\tau = 0$~s, that corresponds to a Newtonian fluid, exhibits the linear relationship between the shear stress and strain rate. Moreover, the Maxwellian liquid of relaxation time $\tau = 0.5 $~s shows a similar behavior, and the ones with $\tau >1$~s show the stress relaxation feature, typical of Maxwell model \cite{Bird}, in which the peak of shear stress is lowered by higher values of the relaxation time.

\begin{figure}[t]
\centering
{\includegraphics[height=5cm]{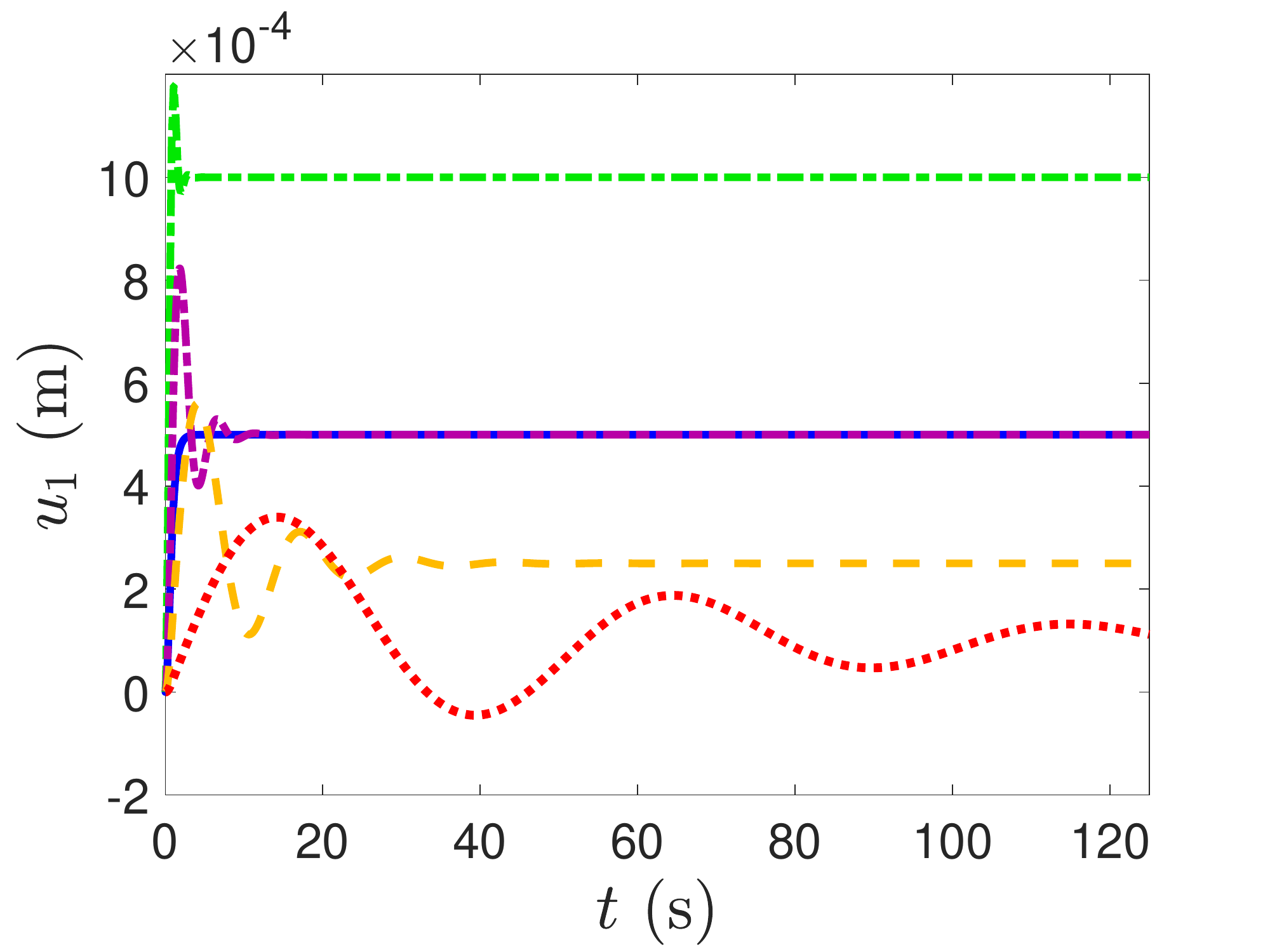}}%
\caption{Evolution of $u_1$, measured from the top-right corner of the sheared membrane, for different values of the relaxation time $\tau = 0$~s (blue solid curve), $0.5$~s (green dashed curve), $1$~s (purple dash-dotted curve), $2$~s (yellow dashed curve), $5$~s (red dotted curve).}\label{fig:U1VsTime}
\end{figure}

Following that, we carry out one last parameter study on the relaxation time in shear flows. For this test case, and different from the previous one in which we imposed an initial velocity for the shear motion, we displace the top boundary of a $2$-element test membrane, by applying a horizontal load $\mb{P}= (0.1, 0)$~N, linearly decreasing in time, with $\mathcal{A}=1$ at $t^{\star}=0$ and $\mathcal{A}=0$ at $t^{\star}=25$. When the load or the deformation is removed, different behaviors occur according to the material model considered. A linearly elastic material bounces back and forth, with no constitutive dissipation. A Newtonian liquid exhibits resistance to the shearing velocity and no elastic behavior. A Maxwell liquid can combine both these two characteristic behaviors, as described in \S~\ref{Sec3}. We measure $u_1$, on the top-right corner of the membrane, and track its evolution in time. In figure \ref{fig:U1VsTime}, we show the values corresponding to the results with $\tau = 0$~s (blue solid curve), $0.5$~s (green dashed curve), $1$~s (purple dash-dotted curve), $2$~s (yellow dashed curve), $5$~s (red dotted curve). We observe how the viscous fluid, corresponding to the curve with $\tau=0$~s, reaches a plateau and does not exhibit any elastic effects. In fact, even when the load is removed, the Newtonian membrane displacement remains constant. On the contrary, the Maxwell liquids exhibit a nearly elastic response in the early times, that is dissipated by viscosity in later times. As stated in the previous paragraph, by increasing the relaxation time $\tau$, at parity of viscosity coefficient, we increase the importance of the viscosity relative to elasticity. In fact, we can see the increasing effects of viscosity in the oscillations with smaller amplitude and larger wavelengths for the curves of $\tau > 1$~s.

\subsection[Membrane deformation under extensional flow]{Membrane deformation under extensional flow}\label{Sec4-3}

\begin{figure}[t]
\centering
\resizebox{.8\textwidth}{!}{
\begin{tikzpicture}
\draw[->] (2.5,2)--(8,2) node[anchor=west] {$Y_1$ (mm)};
\draw[->] (2.5,2)--(2.5,3.2) node[anchor=east] {$T$ (K)};
\draw (2.3,2.2) node [anchor=east] {300};
\draw (2.45,2.2) -- (2.55,2.2);
\draw (2.3,2.65) node [anchor=east] {400};
\draw (2.45,2.65) -- (2.55,2.65);
\draw (2.75,2.2) -- (4,2.2);
\draw (4,1.9) node [anchor=north] {40};
\draw (4,1.95) -- (4,2.05);
\draw (4,2.2) -- (4.5,2.65);
\draw (4.5,1.9) node [anchor=north] {45};
\draw (4.5,1.95) -- (4.5,2.05);
\draw (4.5,2.65) -- (5.5,2.65);
\draw (5.5,1.9) node [anchor=north] {55};
\draw (5.5,1.95) -- (5.5,2.05);
\draw (5.5,2.65) -- (6,2.2);
\draw (6,1.9) node [anchor=north] {60};
\draw (6,1.95) -- (6,2.05);
\draw (6,2.2) -- (7.25,2.2);

\draw (0,0) -- (10,0);
\draw (0,-0.2) -- (10,-0.2);
\draw (0,-0.25) -- (0,-0.15);
\draw (10,-0.25) -- (10,-0.15);
\draw (5,-.4) node[anchor=north] {100 mm};
\draw (10,0) -- (10,1);
\draw (10,1) -- (0,1);
\draw (0,1) -- (0,0);

\draw (-0.2,.5) node[anchor=east] {1 mm};
\draw (-.2,0) -- (-.2,1);
\draw (-.15,0) -- (-.25,0);
\draw (-.15,1) -- (-.25,1);

\draw[->] (10,.5)--(10.25,.5) node [anchor=west]{$\mathbf{v_d}$};
\draw[->] (0,.5)--(0.25,.5) node [anchor=west]{$\mathbf{v_f}$};
\draw[->] (0,0)--(11,0) node[anchor=west] {$Y_1$};
\draw[->] (0,0)--(0,2) node[anchor=east] {$Y_2$};
\end{tikzpicture}%
}
\caption{Schematic of the drawing process of a thin viscoelastic membrane (not in scale), and the temperature profile at the location of the furnace.}\label{fig:DrawingThread}
\end{figure}
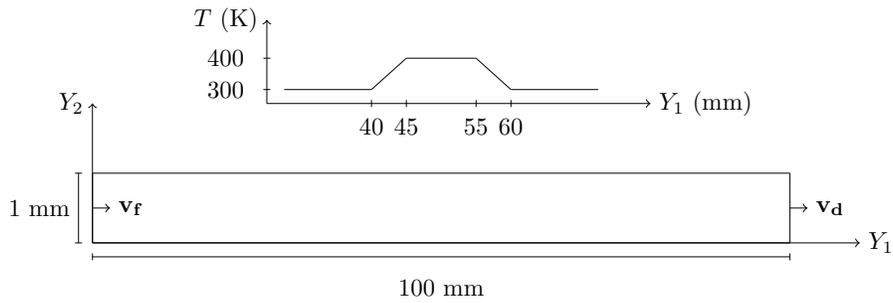

Finally, we consider the application of the drawing of viscoelastic membranes with constant thickness, as a planar study of a more general redrawing process of viscoelastic flat sheets \cite{SrinivasanWeiMahadevan}. Drawing or redrawing processes are manufacturing practices for which a sheet, usually of glass or metal, is heated and stretched to obtain a reduced cross sectional area, such as in the production of glass fibers (see \cite{oKielyBrewardGriffiths2015,Taroni2013,SrinivasanWeiMahadevan} and references therein). We model the sheet as a slender membrane of initial length $L=100$~mm and width $W=1$~mm, with its bottom-left corner coinciding with the origin of the surface coordinate system (see the Appendix) in reference state, $(Y_1,Y_2)$, as depicted in figure \ref{fig:DrawingThread}. The membrane is clamped on its right and left boundaries to rigid walls that move with a drawing velocity on the right boundary, $\mathbf{v_d}=(10^{-3},0)$~m/s, and a feed velocity on the left boundary, $\mathbf{v_f}=(10^{-4},0)$~m/s, respectively. The top and bottom boundary satisfy no-flux and traction-free boundary conditions. As the membrane is drawn, it passes a heated region, representing an idealized furnace, starting at location $Y_1=40$~mm. The furnace temperature follows a linear profile that increases from the ambient temperature, $T_a=300$~K, reaching its maximum, $T_f = 400$~K, that is held constant for \mbox{$45$~mm $< Y_1 < 55$~mm}, and returns to the ambient temperature at $Y_1 = 60$~mm, as shown in figure \ref{fig:DrawingThread}. According to the industrial application of interest, the dimension of the membrane and the furnace can vary \cite{oKielyBrewardGriffiths2015}. We are interested in industrially relevant processes where the furnace zone is short relative to the membrane length, but large relative to the membrane width. Consistent with Srinivasan et al.~\cite{SrinivasanWeiMahadevan}, we assume that the temperature irradiation between the heating device and the viscoelastic membrane is in equilibrium, so that the temperature in the fluid equals the one prescribed by the furnace. As the temperature reaches its maximum, we model the viscosity as linearly dependent on the temperature $T$, according to the following expression
\begin{align}\label{viscosity-tempRelation}
\eta = \eta_a - \frac{\eta_f - \eta_a}{T_f - T_a} (T - T_a) \, ,
\end{align}

\noi where we have considered the difference between the viscosity of the liquid in the furnace, $\eta_f$, and in the ambient, $\eta_a$, to be modeled as $\eta_f - \eta_a = \eta_a /2$, with $\eta_a = 1$~Pa~s.

\begin{figure}[t]
\centering
{\includegraphics[width=7.9cm,valign=t,trim=0in 0.25in 0in 0.278in,clip=true]{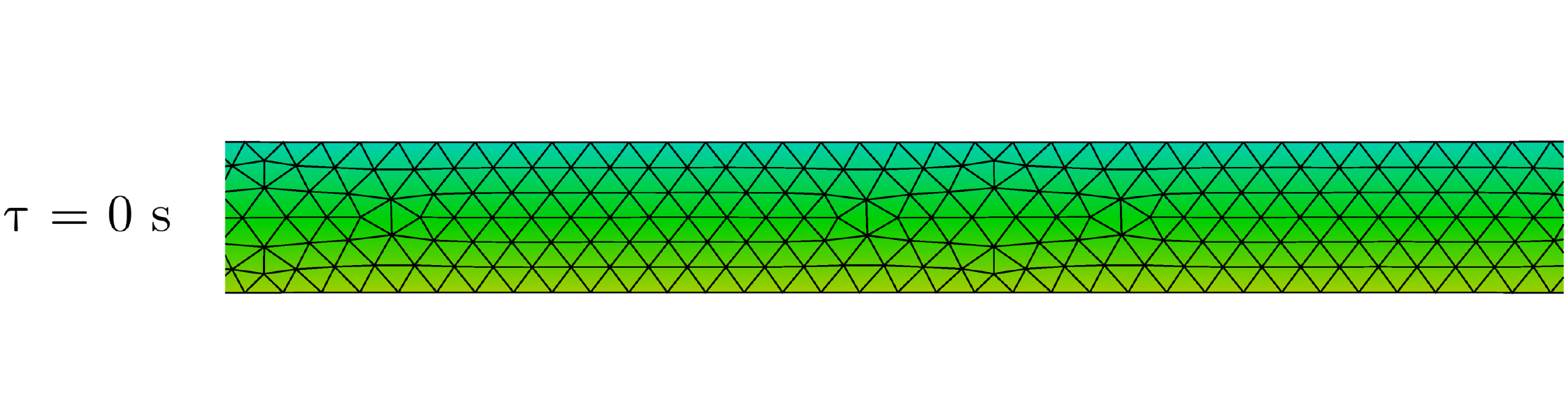}}
{\includegraphics[width=6.9cm,valign=t,trim=0in 0.25in 0in 0.285in,clip=true]{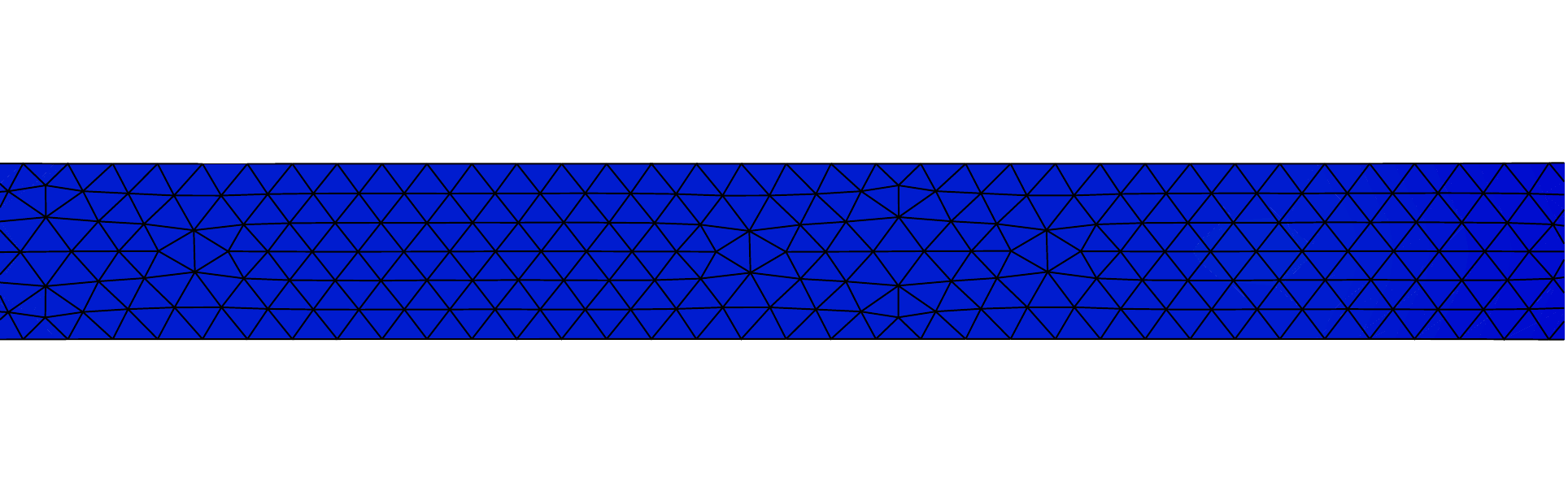}}\\
{\includegraphics[width=7.9cm,valign=t,trim=0in 0.25in 0in 0.278in,clip=true]{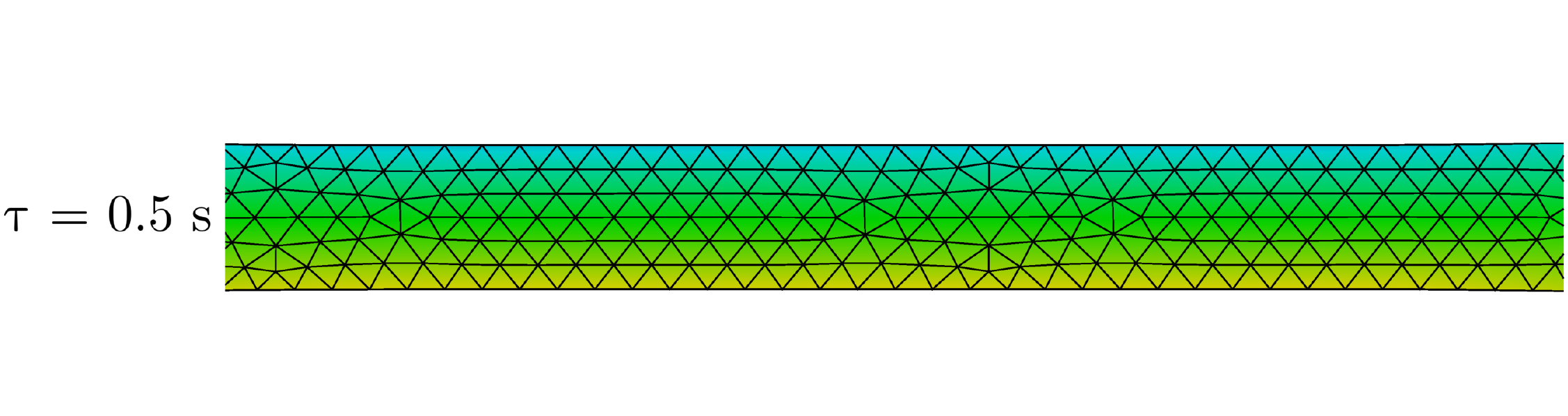}}
{\includegraphics[width=6.9cm,valign=t,trim=0in 0.25in 0in 0.285in,clip=true]{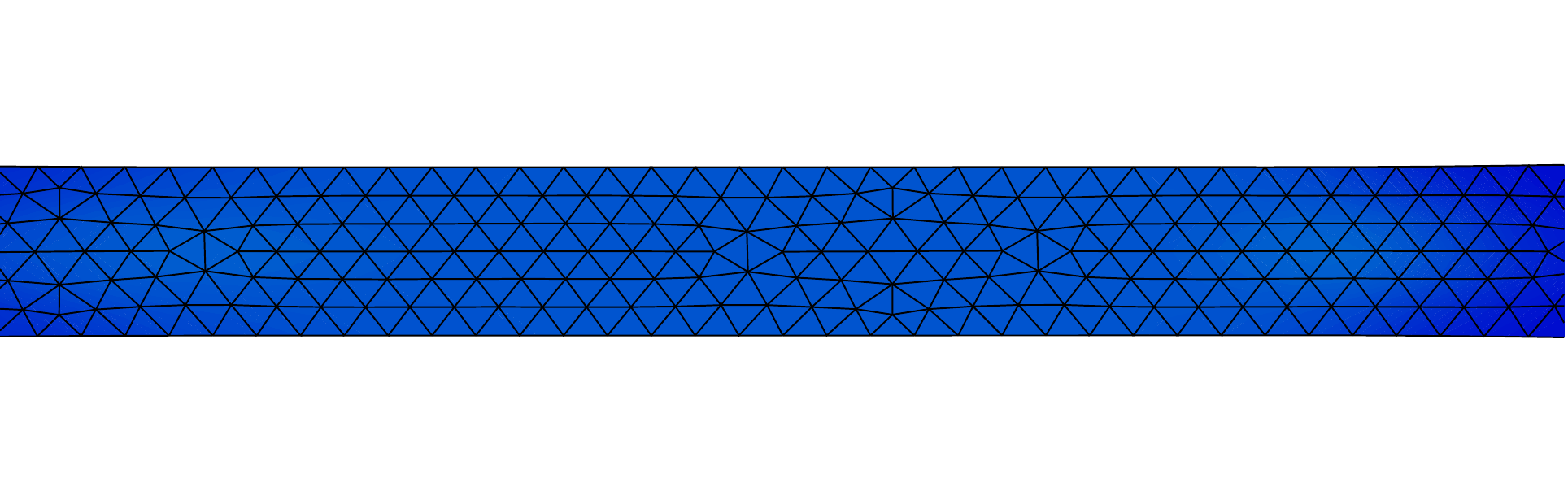}}\\
{\includegraphics[width=7.9cm,valign=t,trim=0in 0.25in 0in 0.278in,clip=true]{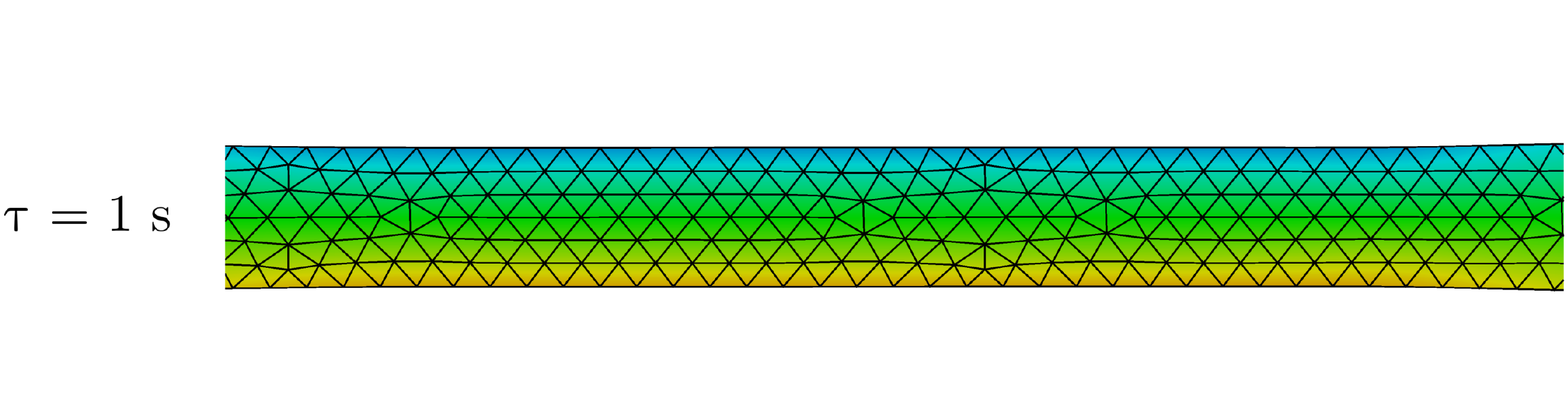}}
{\includegraphics[width=6.9cm,valign=t,trim=0in 0.25in 0in 0.285in,clip=true]{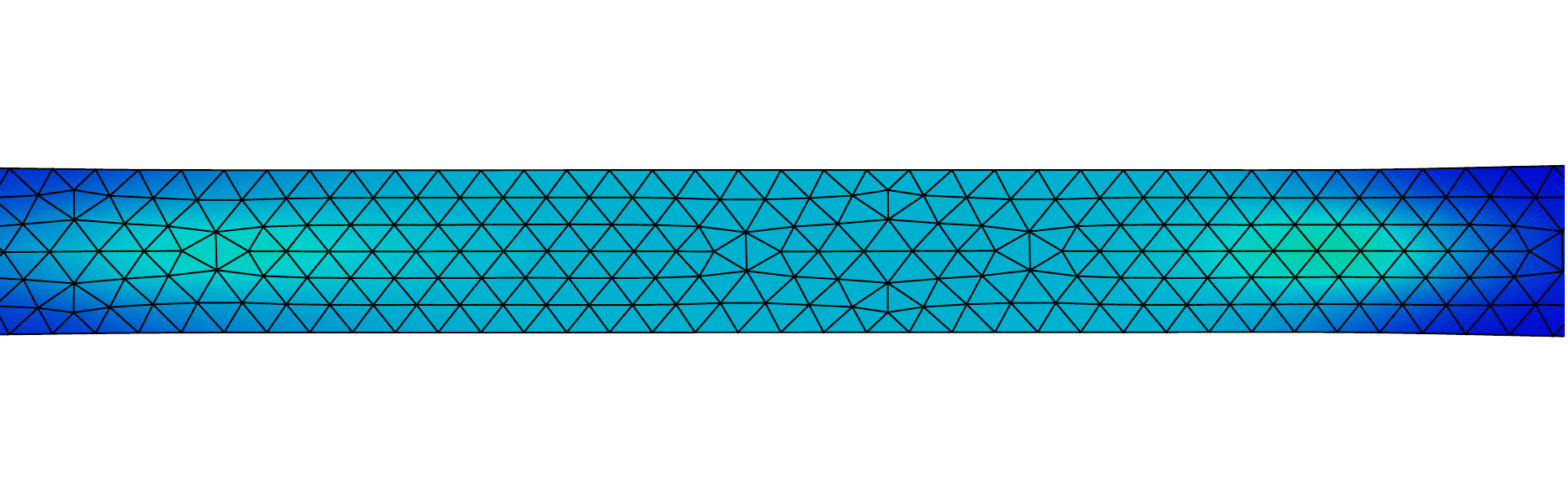}}\\
{\includegraphics[width=7.9cm,valign=t,trim=0in 0.25in 0in 0.278in,clip=true]{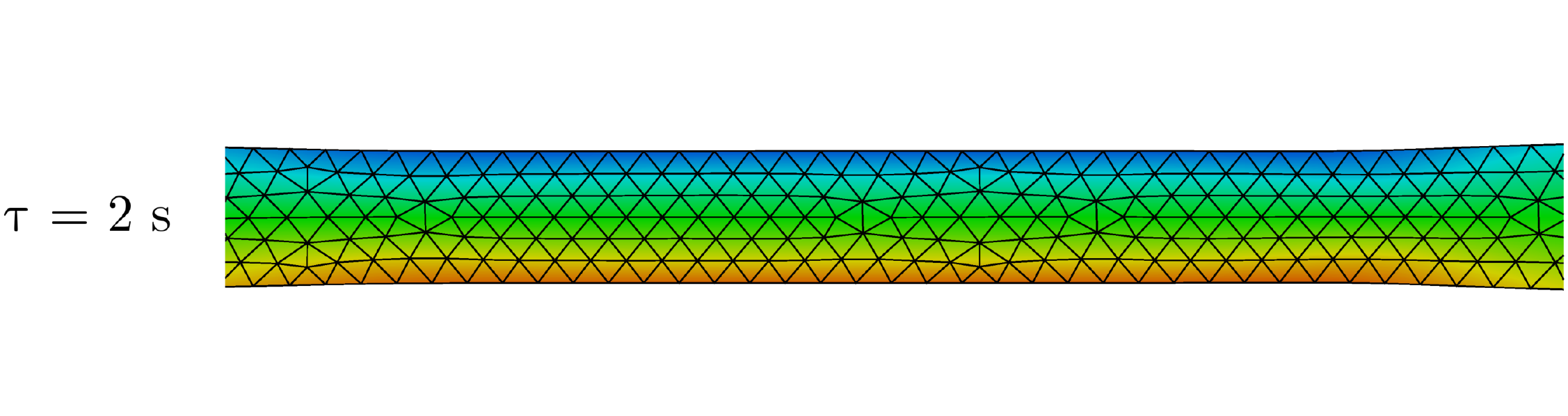}}
{\includegraphics[width=6.9cm,valign=t,trim=0in 0.25in 0in 0.285in,clip=true]{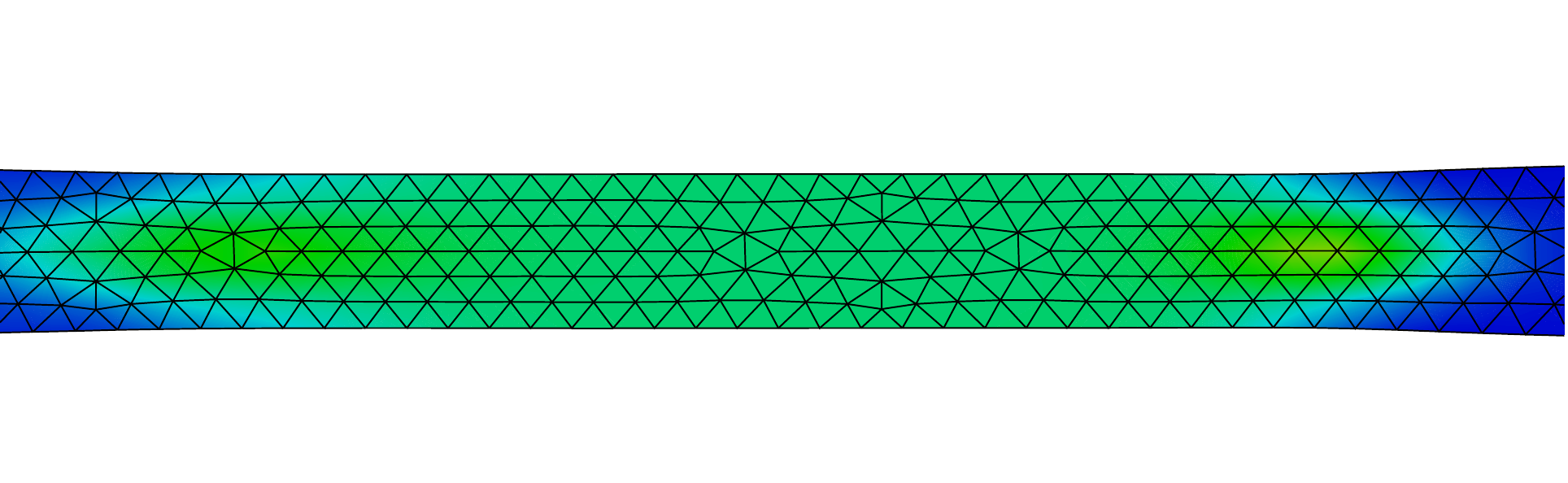}}\\
{\includegraphics[width=7.9cm,valign=t,trim=0in 0.25in 0in 0.278in,clip=true]{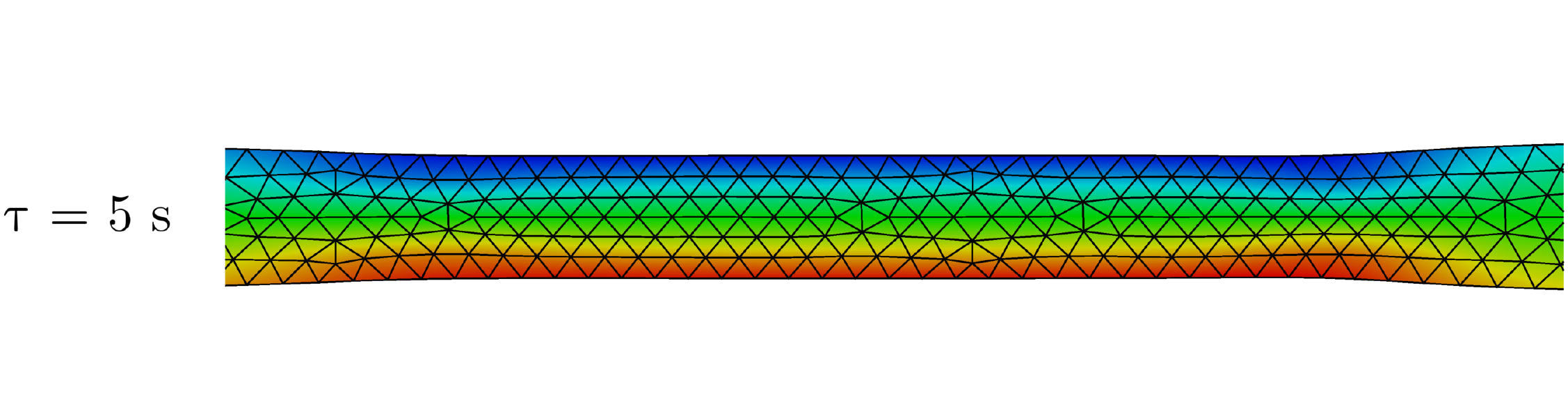}}
{\includegraphics[width=6.9cm,valign=t,trim=0in 0.25in 0in 0.285in,clip=true]{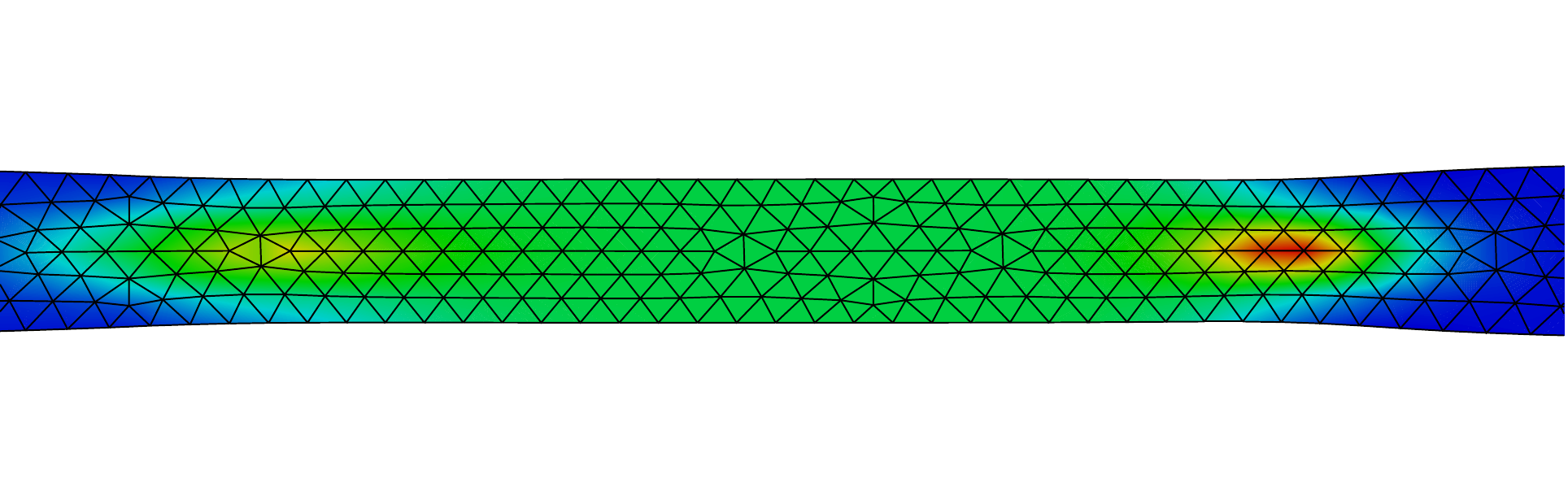}}\\
{\includegraphics[height=2.45cm,valign=t,trim=0cm 0cm 0.9cm 0.7cm]{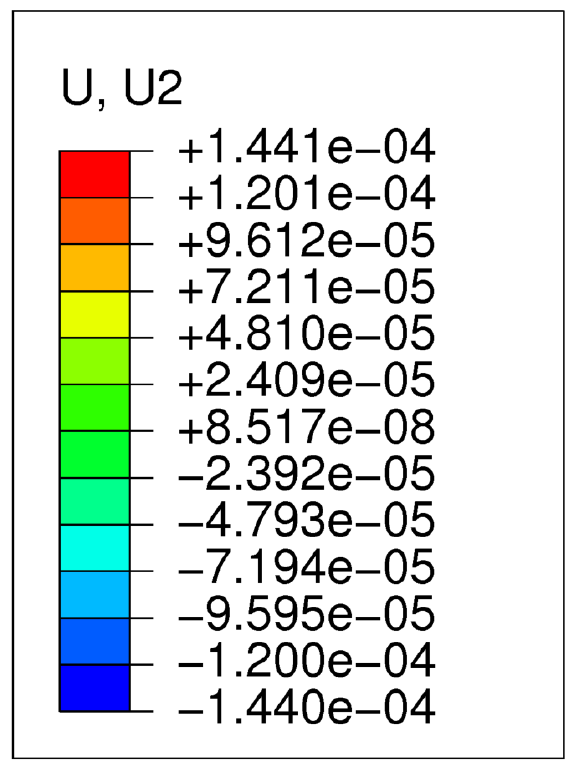}}\hspace{5cm}
{\includegraphics[height=2.45cm,valign=t,trim=0cm 0cm 0.9cm 0.7cm]{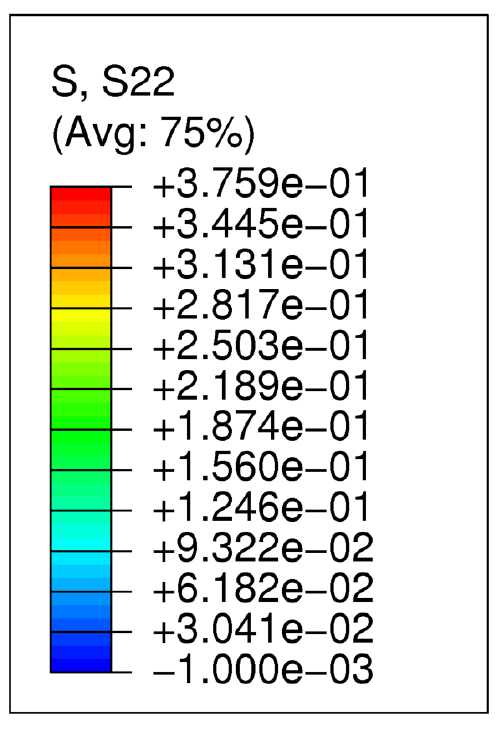}}
\caption{Central region of drawn viscoelastic membranes. On the left panel, contour plots of $u_2$, for the quasi-static solution of drawn membranes, of relaxation time $\tau=0,0.5,1,2,5$~s, (from top to bottom), at $t\sim10$~s. The blue color on the top of the necking region represents the minimum value of $u_2$, ${u_2}_{min}=-1.440\times 10^{-4}$~m, and the red color at the bottom of the necking region represents its maximum value, ${u_2}_{max}\sim-{u_2}_{min}$. On the right panel, contour plots of the second normal stress component, $\sigma_{22}$, at time $t\sim10$~s. The normal stress $\sigma_{22}$ has reached its maximum value, ${\sigma_{22}}_{max}=3.759\times 10^{-1}$~Pa, represented by the red shades, and its minimum value, ${\sigma_{22}}_{min} = -1 \times 10^{-3}$~Pa, represented by the blue shades. The region of maximum stress represents the onset of buckling.}\label{fig:StretchingU2}
\end{figure}

In figure \ref{fig:StretchingU2}, we show contour plots of $u_2$ (on the left panel) and the second normal stress component, $\sigma_{22}$ (on the right panel), for the quasi-static solution of the central region of drawn membranes, of relaxation time $\tau=0,0.5,1,2,5$~s, (from top to bottom), at $t\sim10$~s. As the membranes are stretched, they exhibit some necking in their central part, corresponding to the region of lowest viscosity, consistently with \cite{PfingstagAudolyBoudaoud}. The blue color on the top of the necking region represents the minimum value of $u_2$, ${u_2}_{min}=-1.440\times 10^{-4}$~m, and the red color at the bottom of the necking region represents its maximum value, ${u_2}_{max}\sim-{u_2}_{min}$. In addition, by analyzing the stresses, we can identify the onset of buckling, leading to wrinkling instabilities, that are known to arise when viscous \cite{PfingstagAudolyBoudaoud,SrinivasanWeiMahadevan} or elastic \cite{CerdaMahadevan,CerdaMahadevan2} sheets are stretched. The normal stress $\sigma_{22}$ has reached its maximum value, ${\sigma_{22}}_{max}=3.759\times 10^{-1}$~Pa, represented by the red shades, and its minimum value, ${\sigma_{22}}_{min} = -1 \times 10^{-3}$~Pa, represented by the blue shades. We note that, in the finite element formulation chosen in this work, the stress components are constant on each element. Furthermore, for visualization, the color map representing the stresses is smoothed (within a default threshold of $75\%$, as displayed in the legend of the right panel of figure \ref{fig:StretchingU2}, where $S$ stands for $\bs{\sigma}$ and, similarly, $S22$ for $\sigma_{22}$). Accordingly, there is no visible distinction between the stress value along the edge of an element and its interior. Moreover, we remark that, although the constitutive model chosen does not explicitly represent effects due to a difference in normal stresses, we believe that the region of maximum stress, observed on the right panel of figure \ref{fig:StretchingU2}, suggests the onset of buckling, {similar to the behavior of stretched rubber observed in the literature (see, e.g., \cite{CerdaMahadevan,CerdaMahadevan2}). In figure \ref{fig:U2VsT}, we show the evolution of the point of maximum necking, at the center of the redrawn Newtonian and Maxwellian sheets. We plot $u_2$, at the midpoint of the top boundary of the stretched film, for $\tau = 0$~s (blue solid curve), $0.1$~s (green dashed curve), $0.25$~s (purple dash-dotted curve), $0.5$~s (yellow dashed curve), $0.75$~s (red dotted curve), $1$~s (black solid curve), $2$~s (magenta dash-dotted curve), $5$~s (orange solid curve), and $10$~s (light blue dashed curve), both in logarithmic scale. We can see that the Maxwellian membranes with higher values of the relaxation time exhibit more necking.

\begin{figure}[t]
\centering
{\includegraphics[height=6.5cm,valign=t,trim=0in 0.1in 0.2in 0.15in,clip=true]{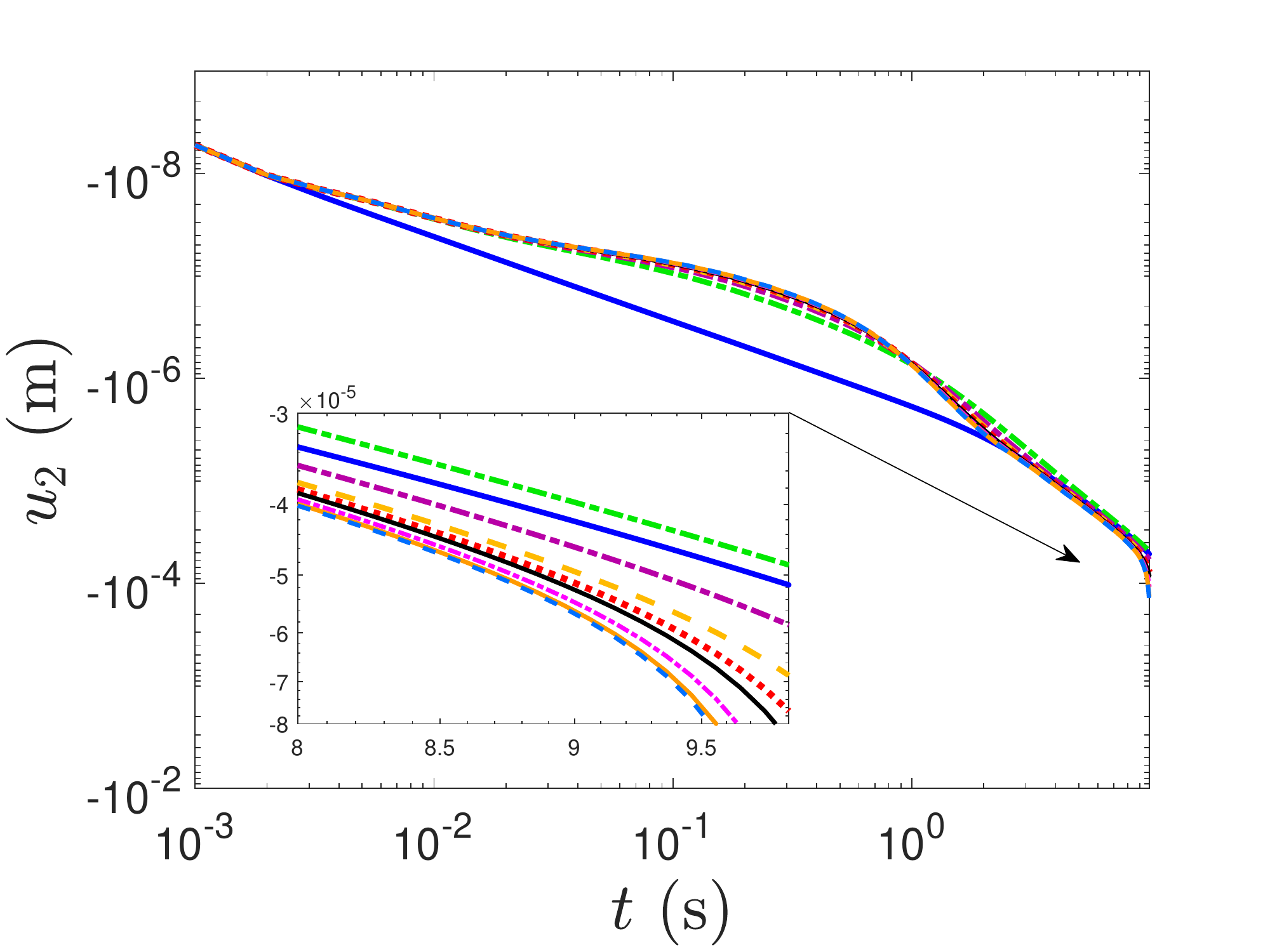}}
\caption{Comparison of $u_2$ at the midpoint of the top boundary of the stretched Newtonian and Maxwellian membranes, for $\tau = 0$~s (blue solid curve), $0.1$~s (green dashed curve), $0.25$~s (purple dash-dotted curve), $0.5$~s (yellow dashed curve), $0.75$~s (red dotted curve), $1$~s (black solid curve), $2$~s (magenta dash-dotted curve), $5$~s (orange solid curve), and $10$~s (light blue dashed curve), both in logarithmic scale. The inset shows a magnification of the graphs for $t\in [8,10]$~s.}\label{fig:U2VsT}
\end{figure}

\begin{figure}[t]
\centering
{\includegraphics[width=8cm,valign=t,trim=0in 0.1in 0.2in 0.15in,clip=true]{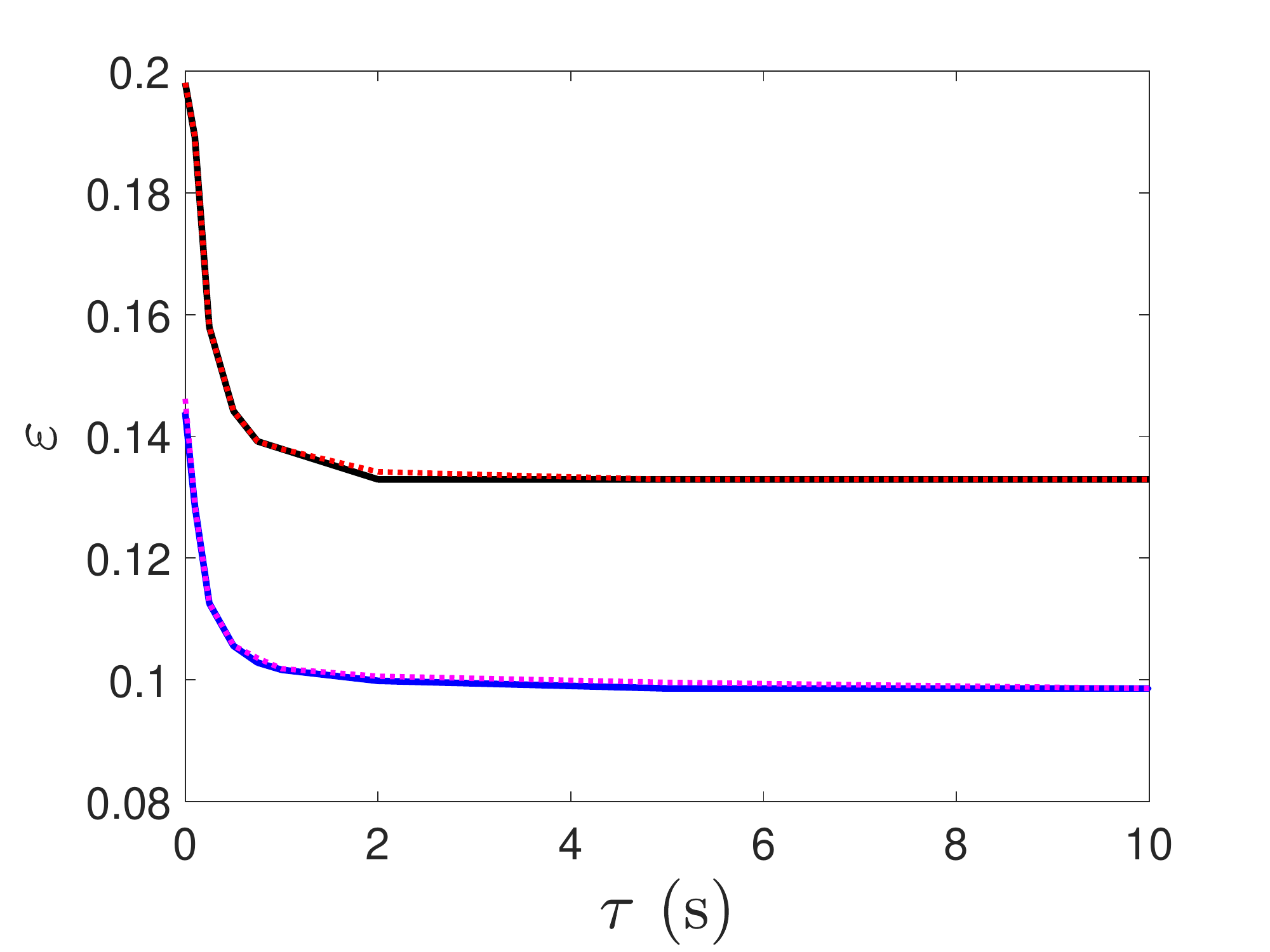}}\\
\caption{Stretch factor, $\varepsilon = {u_1}_{max} /L$, versus the relaxation time $\tau$, for four different sets of feeding and drawing velocities: $\mathbf{v_f} = (10^{-4},0)$~m/s and $\mathbf{v_d} = (10^{-3},0)$~m/s (blue solid curve), $\mathbf{v_f} = (5 \times 10^{-4},0)$~m/s and $\mathbf{v_d} = (10^{-3},0)$~m/s (magenta dotted curve), $\mathbf{v_f} = (10^{-4},0)$~m/s and $\mathbf{v_d} = (5 \times 10^{-3},0)$~m/s (black solid curve), and $\mathbf{v_f} = (5 \times 10^{-4},0)$~m/s and $\mathbf{v_d} = (5 \times 10^{-3},0)$~m/s (red dotted curve).}\label{fig:StretchVsLambda}
\end{figure}

Finally, we investigate the maximum stretch, defined as $\varepsilon = {u_1}_{max} / L$, attained by the elongated membranes before the onset of buckling. This quantity, industrially relevant, can help manufacturers avoid undesired wrinkling instabilities. In figure \ref{fig:StretchVsLambda}, we investigate the influence of the relaxation time, for $\tau \in [0,10]$~s, on $\varepsilon$, for four different sets of feeding and drawing velocities: $\mathbf{v_f} = (10^{-4},0)$~m/s and $\mathbf{v_d} = (10^{-3},0)$~m/s (blue solid curve), $\mathbf{v_f} = (5 \times 10^{-4},0)$~m/s and $\mathbf{v_d} = (10^{-3},0)$~m/s (magenta dotted curve), $\mathbf{v_f} = (10^{-4},0)$~m/s and $\mathbf{v_d} = (5 \times 10^{-3},0)$~m/s (black solid curve), and $\mathbf{v_f} = (5 \times 10^{-4},0)$~m/s and $\mathbf{v_d} = (5 \times 10^{-3},0)$~m/s (red dotted curve). We can see that membranes that are drawn at higher speeds, i.e.~with $\mathbf{v_d} = (5 \times 10^{-3},0)$~m/s, reach a maximum elongation of $20\%$ from their initial length. We moreover notice that membranes with equal drawing velocities exhibit a similar behavior, although the ratio of the magnitude of their drawing to feed velocities, is different, ranging from $10$ for the first and third set of data, to $50$ for the second and fourth ones.

\section[Conclusions]{Conclusions}\label{Sec5}

We have performed a novel numerical investigation of the dynamics of nearly incompressible viscoelastic fluid membranes. We have introduced a displacement-based finite element formulation, in which the stresses are expressed for both viscoelastic fluids of Maxwell type, and viscous (Newtonian) fluids. For the nearly incompressibility condition of both the Newtonian and Maxwellian cases, we have introduced a penalty function, in which the penalty constant is proportional to the viscosity of the fluid. We have validated our numerical implementation with several numerical experiments, demonstrating mesh-independence of our results, and validity of the formulation for near incompressibility, in the limit of the dimensionless parameter $\eta/\widehat{K}$.

We have focused on two main applications of our general numerical framework: shear flow \cite{Bird} and extensional flow in drawing processes \cite{oKielyBrewardGriffiths2015,Taroni2013}. For the case of the simple shear flow of membranes between parallel plates, we have investigated the effect of the relaxation time on the stress relaxation, feature typical of Maxwell liquids \cite{Bird}, and the dynamics. Comparing the behavior of sheared Newtonian and Maxwellian membranes, we have observed the effects of viscoelasticity on the nature of the dynamics, as well as on their final configuration. We have found that Maxwellian membranes deform the most, compared to Newtonian ones, when they are continuously sheared. While they exhibit an elastic response, that is constitutively damped by viscosity, in the case of loading/unloading forcing.

For the drawing process of Newtonian and Maxwellian membranes, with a temperature-dependent viscosity, we have investigated how viscoelasticity affects the necking of the membranes in extensional flows. We have found that higher values of the relaxation time enhance the necking of the stretched membranes. Finally, we have investigated the influence of the relaxation time on the maximum stretch attained by the membranes before the onset of wrinkling instabilities, that are known to arise when viscous \cite{PfingstagAudolyBoudaoud,SrinivasanWeiMahadevan} or elastic \cite{CerdaMahadevan,CerdaMahadevan2} sheets are stretched. We have found that higher values of the relaxation time facilitate the onset of buckling and therefore the emergence of the wrinkling instabilities.

\numberwithin{equation}{section}
\section*{Appendix}
\renewcommand{\theequation}{A\arabic{equation}}
\setcounter{equation}{0}  

We describe here the details of the spatial discretization for each term in equation (\ref{Eq:25}). By linear interpolation, we can specify a position in the triangular element by ${\mathbf{X}} = \xi_\alpha \widetilde{\mathbf{X}}^\alpha$ in the reference configuration, and ${\mathbf{x}} = \xi_\alpha \widetilde{\mathbf{x}}^\alpha$ in the current one. Where $\xi_\alpha$ represents the natural area coordinates, or barycentric coordinates \cite{ZienkiewiczTaylorZhu2013}, such that
\begin{align}\label{Eq:4}
\xi_1 + \xi_2 + \xi_3 &= 1 \, .
\end{align}

\noi Following \cite{TaylorEtAl2005}, to describe the in-plane deformation and stresses of the membrane, it is convenient to introduce a surface coordinate system that lays on the plane of the triangle, denoted by $Y_1$ and $Y_2$, with normal direction $N$ in the reference configuration, and $y_1$, $y_2$ with normal direction $n$ in the current state (see figure \ref{fig:SurfCoordSys}).

In the surface coordinate system, the origin of the coordinates, $(Y_1OY_2)$ and $(y_1oy_2)$ are placed at the nodal locations, $\widetilde{X}^1$ and $\widetilde{x}^1$, respectively. The unit base vectors then may be constructed from the linear displacement triangle, constituted by the three vertices labeled by ($1,2,3$), by aligning the first base vector along the $1$-$2$ side. For simplicity, we denote the edge vectors of the reference configuration by $\mathbf{E}_{12} = \widetilde{\mathbf{X}}^2 - \widetilde{\mathbf{X}}^1$, $\mathbf{E}_{13} = \widetilde{\mathbf{X}}^3 - \widetilde{\mathbf{X}}^1$, $\mathbf{E}_{23} = \widetilde{\mathbf{X}}^3 - \widetilde{\mathbf{X}}^2$, and $\mathbf{e}_{12} = \widetilde{\mathbf{x}}^2 - \widetilde{\mathbf{x}}^1$, $\mathbf{e}_{13} = \widetilde{\mathbf{x}}^3 - \widetilde{\mathbf{x}}^1$, and $\mathbf{e}_{23} = \widetilde{\mathbf{x}}^3 - \widetilde{\mathbf{x}}^2$ for the current configuration. Hence, we define the first unit base vector as
\begin{align}
\mathbf{\hat{e}}_{1}= \frac{\mathbf{e}_{12} }{\| \mathbf{e}_{12}   \|} \, .
\end{align}

\noindent A vector normal to the plane of the triangle is found by $\mathbf{{E}}_{3} = \mathbf{E}_{12} \times \mathbf{E}_{13}$ in the reference state, and $\mathbf{{e}}_{3} = \mathbf{e}_{12} \times \mathbf{e}_{13}$ in the current state. The normal vector in the current state is normalized by
\begin{align}
\mathbf{n} \coloneqq \mathbf{\hat{e}}_{3} = \frac{\mathbf{e}_{3} }{\| \mathbf{e}_{3}   \|} \, ,
\end{align}

\noindent and similarly for the reference state, $\mathbf{N} \coloneqq \mathbf{\widehat{E}}_{3} = {\mathbf{E}_{3} } / {\| \mathbf{E}_{3} \|}$. The second base vector is found by $\mathbf{{E}}_{2} = \mathbf{N} \times \mathbf{E}_{1}$, and analogously by $\mathbf{{e}}_{2} = \mathbf{n} \times \mathbf{e}_{1}$ for the current configuration. Their normalized unit vectors are found, similarly, as $\mathbf{\widehat{E}}_{2} = \mathbf{N} \times \mathbf{\widehat{E}}_{1}$, and $\mathbf{\hat{e}}_{2} = \mathbf{n} \times \mathbf{\hat{e}}_{1}$.

With the base vectors defined above for the plane of the triangle, we can define positions directly as
\begin{align}\label{Eq:9}
y^i= (\mathbf{x} - \widetilde{\mathbf{x}}^1) \cdot \mathbf{\hat{e}}_i \, .
\end{align}

\noi From equation (\ref{Eq:9}), we note that for $\widetilde{\mathbf{y}}^1$, the expression is $\widetilde{y}^1=(\mathbf{\widetilde{x}}^1 - \widetilde{\mathbf{x}}^1) \cdot \mathbf{\hat{e}}_i = 0$. Hence, any position $\mathbf{y}$, found by interpolation of the surface coordinates reduces to
\begin{align}\label{Eq:11}
\mathbf{y} = \xi_\alpha \widetilde{\mathbf{y}}^\alpha = \xi_2 \widetilde{\mathbf{y}}^2 + \xi_3 \widetilde{\mathbf{y}}^3 \, ,
\end{align}

\noi where we have used the summation convention, and equation (\ref{Eq:4}) becomes redundant.

\noi We define the deformation gradient tensor as
\begin{align}\label{Eq:12}
\mathbf{F} = \frac{\partial \mathbf{y}}{\partial \mb{Y}} = \mathbf{I} + \frac{\partial \mathbf{u} }{\partial \mathbf{Y}} \, ,
\end{align}

\noi Moreover, we can write
\begin{align}\label{Eq:13}
\mathbf{F} \frac{\partial \mathbf{Y}}{\partial \bs{\xi}} = \frac{\partial \mathbf{y} }{\partial \mathbf{Y}}\frac{\partial \mathbf{Y}}{\partial \bs{\xi}} = \frac{\partial \mathbf{y}}{\partial \bs{\xi}} \, .
\end{align}

\noi If we denote by $\mb{J}$ the Jacobian transformation tensor for the reference state, and by $\mathbf{j}$ the Jacobian transformation tensor for the current state, we have
\begin{align}\label{Eq:14}
\mathbf{J } = \frac{\partial \mathbf{Y}}{\partial \bs{\xi}} \, , \qquad \mathbf{j } = \frac{\partial \mathbf{y}}{\partial \bs{\xi}} \, .
\end{align}

\noi Hence, we can express the deformation gradient as
\begin{align}\label{Eq:15}
\mathbf{F} = \mathbf{j}\mathbf{G}\, ,
\end{align}

\noi where we have used $\mathbf{G}=\mathbf{J}^{-1}$. Following closely the derivation by Taylor et al.~in \cite{TaylorEtAl2005}, we can expand the expressions for the matrices $\mb{J}$ and $\mb{j}$, by taking into considerations that $\mathbf{E}_{12}$ is orthogonal to the unit vector $\mathbf{\widehat{E}}_{1}$, and analogously $\mathbf{e}_{12}$ is orthogonal to the unit vector $\mathbf{\hat{e}}_{1}$, they become
\begin{align}\label{Eq:18}
\mathbf{J} =
\left[
\begin{array}{cc}
\| \mb{E}_{12} \| & {\mb{E}_{12}^T \mb{E}_{13}}/{\| \mb{E}_{12} \| } \\
0                 & {\mathbf{\widehat{E}}_{3}}/{\| \mb{E}_{12} \|}
\end{array}
\right]\, ,
\end{align}
\noi and
\begin{align}\label{Eq:19}
\mathbf{j} =
\left[
\begin{array}{cc}
\| \mb{e}_{12} \| & {\mb{e}_{12}^T \mb{e}_{13}}/{\| \mb{e}_{12} \| } \\
0                 & {\mathbf{\widehat{e}}_{3}}/{\| \mb{e}_{12} \|}
\end{array}
\right]\, .
\end{align}

\noi We note that the symmetric part of the displacement gradient is defined as ${H}_{ij} = \partial u_i / \partial x_j $, and can be recast as
\begin{align}\label{Eq:H}
\mathbf{H} =  \mathbf{F} - \mathbf{I} \, .
\end{align}

\noi Thus,
\begin{align}\label{Eq:epsilon}
\bs{\epsilon} = \frac{1}{2} \left( \mathbf{H} + \mb{H}^T \right) \, .
\end{align}

\noi We can then define
\begin{align}\label{Eq:20}
\mathbf{C} = \mathbf{F}^T\mathbf{F} =  \mathbf{J}^{-T} \mathbf{j}^T  \mathbf{j}  \mathbf{J}^{-1} =  \mathbf{G}^T  \mathbf{g}  \mathbf{G}\, ,
\end{align}

\noi where we have used $\mathbf{g}= \mathbf{j}^T  \mathbf{j}$. We rewrite equation (\ref{Eq:20}) in component form as
\begin{align}\label{Eq:27}
C_{IJ} = G_{iI} g_{ij} G_{jJ}\, , \quad \textrm{ with } i,j = 1,2 \; , \textrm{ and } I,J = 1,2 \, ,
\end{align}

\noi where the components of the matrix $ \mathbf{G}$ are
\begin{align}\label{Eq:28}
 G_{11} = \frac{1}{J_{11}} \; , G_{22} = \frac{1}{J_{22}} \; ,G_{12} = \frac{- J_{12}}{J_{11}J_{22}} \; , G_{21} = 0 \, .
\end{align}

\noi We can now find the relations among the indices needed for the term $\delta \bs{\epsilon} ^T \bs{\sigma} $ in equation (\ref{Eq:25}), first by noting that
\begin{align}\label{Eq:29}
\delta C_{IJ} \sigma_{IJ} = G_{iI} \delta{g}_{ij} G_{jJ} \sigma_{IJ} = \delta g_{ij} s_{ij} \, ,
\end{align}

\noi where the variable $s_{ij}$, related to stress, is defined by
\begin{align}\label{Eq:30}
s_{ij} = G_{iI} G_{jJ} \sigma_{IJ} \, .
\end{align}

\noi We can rewrite the last transformation in matrix form
\begin{align}\label{Eq:31}
s_{ij} = \mb{Q}^T \bs{\sigma} \, ,
\end{align}

\noi where $\mb{Q}$ is a matrix of the change of index, defined by
\begin{align}\label{Eq:32}
\mb{Q} = \left[
\begin{array}{ccc}
G_{11}^2 & 0 & 0\\
G_{12}^2 & G_{22}^2 & G_{12}G_{22}\\
2 G_{11}G_{12} & 0 & G_{11}G_{22}
\end{array}
\right] \, .
\end{align}

\noi We can use equation (\ref{Eq:29}) and write the second term on the right-hand side of equation (\ref{Eq:25}), as
\begin{align}\label{Eq:33}
\int_{\Omega^{(e)}} \delta \bs{\epsilon}^T \bs{\sigma} h \, d\, A = \int_{\Omega^{(e)}} \frac{h}{2}\delta  C_{IJ} \sigma_{IJ} d\,A  =  \frac{h}{2} \delta g_{ij} s_{ij} A^{(e)}\, ,
\end{align}

\noi where the area of a triangular element in the reference configuration, $A^{(e)}$, can be calculated given any two vectors on the reference configuration triangle, e.g.~$\mathbf{E}_{12} $, and $\mathbf{E}_{13}$, by $A^{(e)} = {\| \mathbf{E}_{12} \times \mathbf{E}_{13}\|}/{ 2 }$. It is convenient to rewrite equation (\ref{Eq:33}) in matrix form
\begin{align}\label{Eq:35}
\frac{1}{2} \delta C_{IJ} S_{IJ} = \left[\delta \epsilon_{11} \; \; \delta \epsilon_{22} \; \; 2 \delta \epsilon_{12} \right]
 \left[
\begin{array}{c}
\sigma_{11}\\
\sigma_{22}\\
\sigma_{12}
\end{array}
\right] = \delta \bs{\epsilon}^T \bs{\sigma} \, ,
\end{align}

\noi or, in terms of the expression found in equation (\ref{Eq:33})
\begin{align}\label{Eq:36}
\frac{1}{2} \delta {g}_{ij} {s}_{ij} = \left[\delta g_{11} \; \; \delta g_{22} \; \; 2 \delta g_{12} \right]
 \left[
\begin{array}{c}
s_{11}\\
s_{22}\\
s_{12}
\end{array}
\right] = \frac{1}{2}\delta \bs{g}^T \bs{s} \, .
\end{align}

\noi We can finally write
\begin{align}\label{Eq:37}
\delta \bs{\epsilon} = \frac{1}{2} \delta \mb{C} = \mb{Q} \mb{b} \delta \tilde{\mb{x}} \, ,
\end{align}

\noi where the vector $\tilde{\mb{x}}$ represents the three nodal values stacked in a ($9 \times 1$) column vector, and $\mb{b}$ is the strain-displacement matrix, given by
\begin{align}\label{Eq:bMat}
\mb{b} =
\left[
\begin{array}{ccc}
-\mathbf{e}_{12}^T & \mathbf{e}_{12}^T  & \mb{0} \\
-\mathbf{e}_{13}^T & \mb{0} & \mathbf{e}_{13}^T  \\
-(\mathbf{e}_{12} + \mathbf{e}_{13})^T & \mathbf{e}_{13}^T & \mathbf{e}_{12}^T
\end{array}
\right]\, .
\end{align}

\noi Finally, we can form the divergence operator matrix, for each element, $\mb{B}^{(e)}$, in equation (\ref{Eq:25}), in terms of variations of the displacement for each element, as
\begin{align}\label{Eq:40}
\mb{B}^{(e)} = \mb{Q} \mb{b} \, .
\end{align}

\noi We next need to define the matrix $\mb{M}^{(e)}$, in equation (\ref{Eq:25}), representing the mass matrix for each element, whose components are given by
\begin{align}\label{MassMatrix}
\mb{M}^{(e)}_{\alpha \beta} =\int_{\Omega^{(e)}} \rho h \xi_{\alpha} \xi_{\beta} d \, A \;\mb{I} \, .
\end{align}

\noi The last vector used in equation (\ref{Eq:25}), $\mb{\widetilde{F}_b}$, represents the constant nodal body force, such as gravity, that is trivially linearly interpolated at the nodes, and its description is omitted here.

\cleardoublepage
\section*{References}
\bibliography{References}

\end{document}